%% file: main.tex
\documentclass[a4paper, 10pt, conference]{IEEEtran}      

\IEEEoverridecommandlockouts                              

\usepackage[utf8]{inputenc}
\usepackage[T1]{fontenc}
\usepackage[utf8]{inputenc}
\usepackage{etex}
\usepackage{amsmath, amssymb, amsfonts, bm, nccmath}                        
\usepackage{booktabs}                       
\usepackage{graphicx}
\usepackage{amssymb, stfloats, amssymb, amsfonts, rotating, acronym}
\usepackage{multirow}
\usepackage{relsize}
\usepackage{array}
\usepackage{tabularx}
\usepackage{tikz}
\usetikzlibrary{patterns}
\usetikzlibrary{arrows,positioning,shapes.geometric,spy,matrix}
\usetikzlibrary{backgrounds}
\usepackage{mathtools}
\usepackage{color}
\usepackage{xcolor}
\usepackage{pgfplots}
\pgfplotsset{compat=newest}
\usepgfplotslibrary{groupplots}
\usepgfplotslibrary{fillbetween}
\usepackage{cite}
\usepackage{adjustbox}
\usepackage{colortbl}
\usepackage{lipsum}
\usepackage[linesnumbered,algoruled,boxed,lined]{algorithm2e}
\usepackage[normalem]{ulem}
\usepackage{cuted}

\newcommand{\fixme}[2]{\ifx&#2&{\leavevmode\color{red}#1}\else{\leavevmode\color{red}FIXME\{}#1{\leavevmode\color{red}\}}\footnote{{\leavevmode\color{red}#2}}\PackageWarning{Fixme}{#1: #2}\fi}

\input{colors.tex}

\title{GRAND-EDGE: A Universal, Jamming-resilient Algorithm with Error-and-Erasure Decoding}

\begin{document}

\author{\IEEEauthorblockN{Furkan Ercan\textsuperscript{$\dagger$}, Kevin Galligan\textsuperscript{*}, David Starobinski\textsuperscript{$\dagger$}, Muriel M{\'e}dard\textsuperscript{$\mathsection$}, Ken R. Duffy\textsuperscript{*}, Rabia Tugce Yazicigil\textsuperscript{$\dagger$}}
\IEEEauthorblockA{\textsuperscript{$\dagger$}Department of Electrical and Computer Engineering, Boston University, Boston, MA, USA \\
\textsuperscript{$\mathsection$}Department of Electrical Engineering and Computer Science, MIT, Cambridge, MA, USA \\
\textsuperscript{*}Hamilton Institute, Maynooth University, Ireland
}}

\maketitle

\thispagestyle{empty}
\pagestyle{empty}

\begin{abstract}
Random jammers that overpower transmitted signals are a practical concern for many wireless communication protocols. As such, wireless receivers must be able to cope with standard channel noise and jamming (intentional or unintentional). To address this challenge, we propose a novel method to augment the resilience of the recent family of universal error-correcting GRAND algorithms. This method, called Erasure Decoding by Gaussian Elimination (EDGE), impacts the syndrome check block and is applicable to any variant of GRAND. We show that the proposed EDGE method naturally reverts to the original syndrome check function in the absence of erasures caused by jamming. We demonstrate this by implementing and evaluating GRAND-EDGE and ORBGRAND-EDGE. Simulation results, using a Random Linear Code (RLC) with a code rate of $105/128$,  show that the EDGE variants lower 
both the Block Error Rate (BLER) and the computational complexity by up to five order of magnitude compared to the original GRAND and ORBGRAND algorithms. We further compare ORBGRAND-EDGE to Ordered Statistics Decoding (OSD), and demonstrate an improvement of up to three orders of magnitude in the BLER. 

\end{abstract}

\section{Introduction}

What do a Wi-Fi router, a Bluetooth speaker, and a leaky microwave oven have in common? They all broadcast wireless signals in the same frequency range. Modern wireless technologies ensure reliable communication by employing techniques, such as subcarrier frequency hopping \cite{Torrieri2018} or cyclic prefixing \cite{tse_book}. Yet, powerful interference, such as a leaky microwave oven operating at a nearby frequency, may disrupt  communication between devices, e.g., by rendering a substantial portion of the subcarrier frequencies unusable. Under such adversarial conditions, although the received signal strength indicator (RSSI) may alert the receiver about channel anomalies \cite{wu_rssi}, the retrieved data that is overpowered by interference is practically lost.  

The purpose of this work is to add resilience against such jamming  events.
We consider a channel model in which a powerful jammer impacts the transmitted data randomly, at a bit-level. In this case, the individual non-jammed bits are subject to the additive white Gaussian noise (AWGN) of the channel, and the jammed bits are impacted by a more extreme, additive noise. We aim to develop an error correction algorithm that provides data recovery capabilities under these adversarial conditions. We propose doing so in conjunction with Guessing Random Additive Noise Decoding (GRAND), a recently proposed error correction decoder algorithm that can work with any codebook \cite{duffy2019grand}. 
Besides the original hard-information GRAND algorithm, soft-information-based variants are also available \cite{solomon2020sgrand, duffy2022srgrand, duffy2021orbgrand}. Among them, the Ordered Reliability Bits GRAND (ORBGRAND) \cite{duffy2022orbgrand} lends itself to practical hardware implementations while maintaining a near maximum-likelihood (ML) decoding performance \cite{abbas2022orbgrand}. 

In this work, we propose a jamming-resilient algorithm based on GRAND algorithm and its variants. Our method seeks to perform error-correction on the non-jammed bits of the received frame, and perform erasure-correction on the jammed bits. It does so by: 
\begin{enumerate}
    \item Identifying jammed bits through RSSI observation;
    \item Performing error correction on non-jammed bits. As jammed bits occur randomly in any part of the received codeword, a challenge is to error-correct a partial code that changes on each communication. This challenge requires a universal decoding approach, for which we use hard- and soft-information variants of the GRAND algorithm;
    \item Having error-corrected the non-jammed bits, determining the values of the jammed bits through Gaussian elimination.
\end{enumerate}
We achieve this capability by empowering the syndrome check function of any GRAND-based algorithm with the ability of restoring the erased bits in a received frame. This upgraded syndrome check method is called \textit{Erasure Decoding by Gaussian Elimination}, or EDGE in short. 

In general, most existing error-and-erasure decoding (EED)
are based on specific decoding schemes. Some EED schemes are designed to retrieve corrupted \emph{frames} rather than \emph{bits} in the presence of erasures.  This includes schemes, such as random linear network coding (RLNC) \cite{koetter2003algebraic,katti2008xors}, product/staircase codes \cite{lukas2021} and fountain codes \cite{fountain2022}. On the other hand, our proposed EDGE method operates at the bit-level, and can be used with \emph{any} linear code.

We introduce two variants of the EDGE subroutine: one with hard-information (GRAND-EDGE) and the other with soft-information (ORBGRAND-EDGE) decoding. Simulation results demonstrate that the EDGE subroutine improves both the block-error rate (BLER) performance and the computational complexity by up to five orders of magnitude, under adversarial channel conditions. We also compare ORBGRAND-EDGE with the Ordered Statistics Decoding (OSD) algorithm \cite{osd1995} which is also based on Gaussian elimination. We show that the proposed ORBGRAND-EDGE algorithm improves the BLER by up to three orders of magnitude while achieving lower complexity compared to OSD.

The rest of the paper is organized as follows. The background on GRAND is detailed  Section~\ref{sec:bg}. In Section~\ref{sec:bg}, the EDGE subroutine and its applications with universal hard- and soft-decoding variants, GRAND-EDGE and ORBGRAND-EDGE, are presented. The benefits of the proposed universal error-and-erasure decoding via simulation results are presented in Section~\ref{sec:res}, followed by concluding remarks in Section~\ref{sec:conc}. 

\section{The GRAND Algorithm}\label{sec:bg}


Guessing Random Additive Noise Decoding (GRAND) \cite{duffy2019grand} is a recently introduced universal algorithm capable of decoding any code, using codebook membership checks. For a received (hard-information) frame vector $\mathbf{r}$, the membership (syndrome) check is performed as
\begin{equation}\label{eq:pc}
    \mathbf{H}\cdot\mathbf{r^{\top}} \text{,}
\end{equation}
where $\mathbf{H}$ is the codebook-specific parity-check matrix of a linear code.
Unlike traditional decoding algorithms, GRAND focuses on the noise component of the received frame. If (\ref{eq:pc}) is not equal to an all-zero vector ($\mathbf{0}$), then the received sequence $\mathbf{r}$ is not a member of the codebook due to noise-corrupted bits. 
The GRAND algorithm trials putative error sequences, represented by $\mathbf{e}$, in maximum likelihood order. It subtracts each of them from $\mathbf{r}$ until it finds one that satisfies
\begin{equation}\label{eq:pc2}
    \mathbf{H}\cdot(\mathbf{r} \oplus \mathbf{e})^{\top} = \mathbf{0} \text{.}
\end{equation}
where $\oplus$ is the modulo-2 sum operator.
\begin{figure}[t]
\includegraphics[width=\columnwidth]{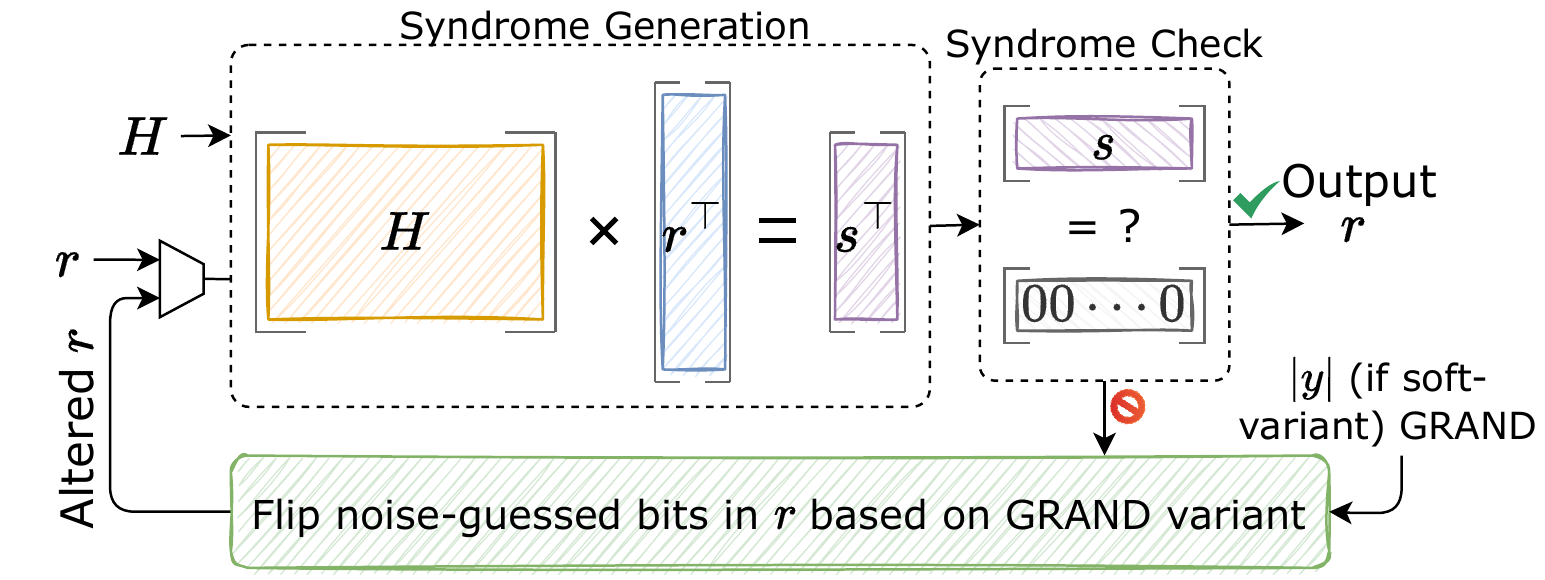}
\caption{A component-level description of the GRAND algorithm family, with in-detail syndrome generation process.}
\label{fig:grand}
\end{figure}


Different ordering of guessing noise sequences lead to different variations of GRAND. A high-level description of the GRAND algorithm family is depicted in Fig.~\ref{fig:grand}. Among the variants, Ordered Reliability Bits GRAND (ORBGRAND)~\cite{duffy2021orbgrand} is a soft-information decoding algorithm that orders the putative noise sequences based on the \textit{logistic weight} ($LW$) of their sorted LLR magnitudes, and is amenable to hardware implementation~\cite{abbas2022orbgrand}. 

\section{The GRAND-EDGE Algorithm}\label{sec:3}

In this Section, the channel model is characterized first. Then the EDGE subroutine is explained in detail, followed by a closer look at the Gaussian elimination process. Finally, the proposed GRAND-EDGE algorithm is described.

\subsection{Channel Model}\label{sec:3:chn}

\begin{figure}[t]
\includegraphics[width=1.02\columnwidth]{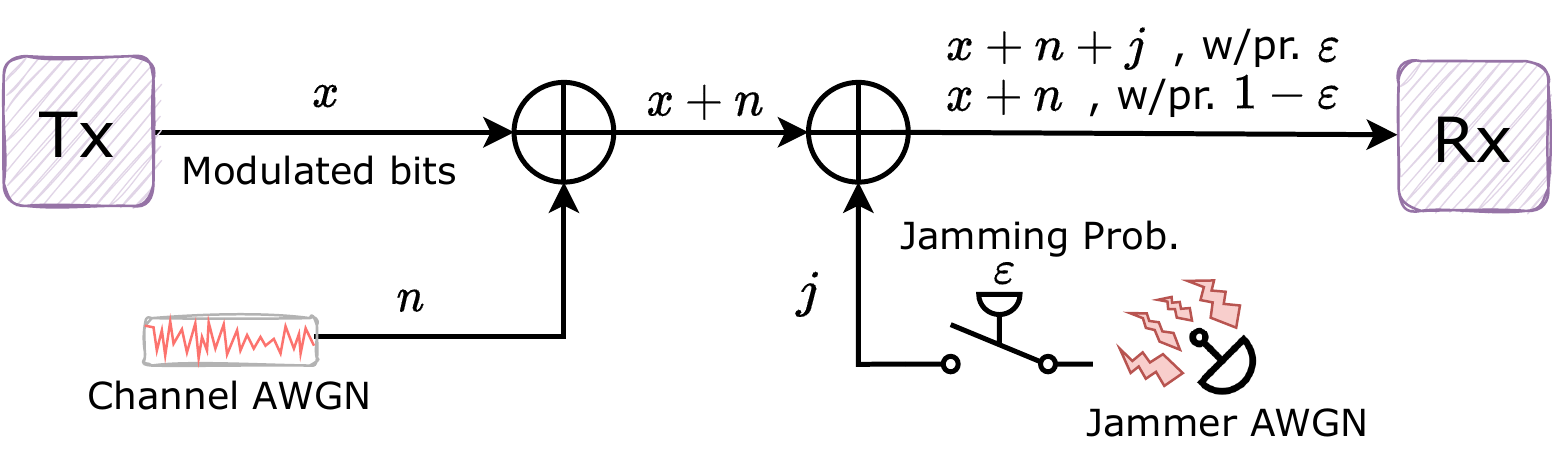}
\caption{The channel model.}
\label{fig:chn}
\end{figure}

In this work, we consider an AWGN channel model that is randomly disrupted by a jammer, as depicted in Fig.~\ref{fig:chn}. The additive channel noise instance, represented by $n$, is added to the modulated signal $x$ carried over a specific frequency. A powerful jammer instance $j$, activated with a probability $\epsilon$ may be added to the transmitted signal. We assume that the probability of jamming $\epsilon$ is on the bit-level rather than the frame-level. Therefore, the received signal $y$ can be expressed as
\begin{equation}\label{eqn:chn}
y=\left\{
  \begin{array}{@{}ll@{}}
    x + n + j & \text{with probability } \epsilon; \\
    x + n & \text{otherwise.}
  \end{array}\right.
\end{equation}
The jammer is also modeled as AWGN, but with a variance that is far greater than that of the channel AWGN. Therefore, if the received signal magnitude is suspiciously stronger than expected, then it is assumed that the signal is jammed and its value is invalidated. 

For frame-level jamming or erasures, such as lost frames due to undecodable preambles, can also be converted to bit-level erasures by simple interleaving techniques, such as in \cite{riaz_icc2022}.

\subsection{The EDGE Subroutine}\label{sec:3:edge}

\begin{figure}[t]
\includegraphics[width=1.05\columnwidth]{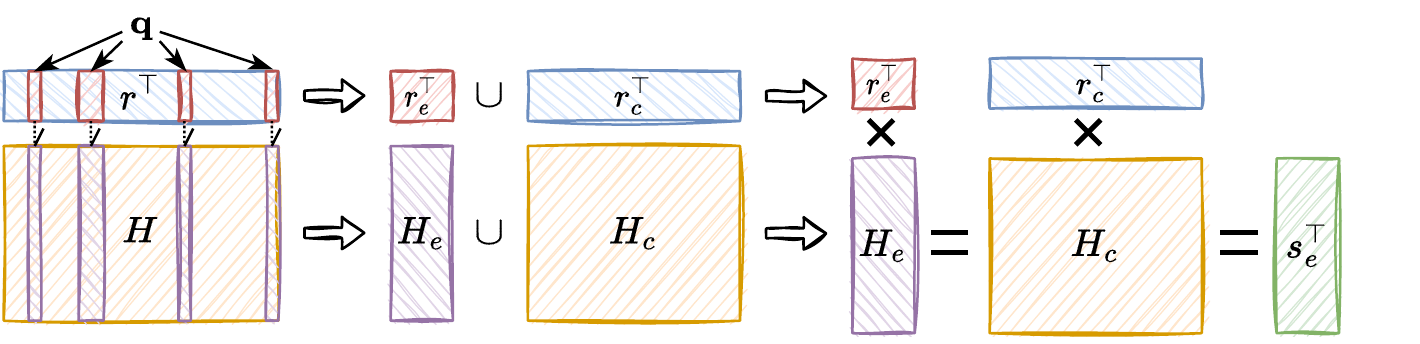}
\caption{Isolation of erasures ($\mathbf{r_e}$) from the received (hard-decision) codeword and corresponding parity-check columns ($\mathbf{H_e}$) from the H-matrix, followed by the calculation of erasure syndrome ($\mathbf{s_e}$).}
\label{fig:separate}
\end{figure}

Let $\mathbf{p} = \{0, 1, \cdots N-1\}$ represent the total set of 0-indexed indices of the received vector of length $N$, and let $\mathbf{q}$ represent the set of imputed (erased) indices in the received vector, such that $\mathbf{q}$ has $e$ elements, $\mathbf{q} \subseteq \mathbf{p}$, and $q_i < q_{i+1}, \forall{i}$. As shown in Fig.~\ref{fig:separate}, given $\mathbf{q}$, let us split the received vector $\mathbf{r}$ and the parity-check matrix $\mathbf{H}$
\begin{equation} \label{eq:split}
\begin{split}
\mathbf{r} & = \mathbf{r_e} \cup \mathbf{r_c} \text{,} \\
\mathbf{H} & = \mathbf{H_e} \cup \mathbf{H_c} \text{,}
\end{split}
\end{equation}
where $\mathbf{r_e}$ and $\mathbf{r_c}$ represent the erased and non-erased subsets of $\mathbf{r}$ with sizes $e$ and $N-e$, respectively. Similarly, $\mathbf{H_e}$ is a $(N-k)\times e$ matrix that contains the columns of $\mathbf{H}$ which correspond to the erased bits, and $\mathbf{H_c}$ contains the remaining columns. Note that the original sets in (\ref{eq:split}) can be reconstructed from the separated subsets, using $\mathbf{q}$. The parity-check equation described in (\ref{eq:pc}) can be expanded as
\begin{equation}\label{eq:pcnew}
    \mathbf{H_e}\cdot\mathbf{r_e^{\top}} \oplus \mathbf{H_c}\cdot\mathbf{r_c^{\top}} = \mathbf{0} \text{.}
\end{equation}
Here, all the components are known on the receiver side except $\mathbf{r_e}$. Let us define the \textit{erasure syndrome}, $\mathbf{s_e}$, as 
\begin{equation}\label{eq:se}
\mathbf{s_e^{\top}} = \mathbf{H_c}\cdot\mathbf{r_c^{\top}} \text{,}
\end{equation}
and transfer it to the right-hand side of (\ref{eq:pcnew}) in the binary domain, to obtain
\begin{equation}\label{eq:pc-era}
    \mathbf{H_e}\cdot\mathbf{r_e^{\top}} = \mathbf{s_e^{\top}} \text{.}
\end{equation}
as visualized in Fig.~\ref{fig:separate}. 

The EDGE subroutine performs Gaussian elimination on the \textit{linear set of equations} in (\ref{eq:pc-era}) to find $\mathbf{r_e}$. To find a unique solution for $\mathbf{r_e}$, the number of equations must not be smaller than the number of variables. In other words, the number of rows of $\mathbf{H_e}$ must be equal to or greater than its columns, therefore, we must first satisfy $e \leq N-k$. 

  

\begin{algorithm}[t]
\caption{EDGE Subroutine Initialization}\label{alg:edge:init}
\SetKwData{Left}{left}\SetKwData{This}{this}\SetKwData{Up}{up}
\SetKwFunction{Union}{Union}\SetKwFunction{FindCompress}{FindCompress}
\SetKwInOut{Inputs}{Inputs}\SetKwInOut{Outputs}{Outputs}
  \Inputs{$\mathbf{r}$, $\mathbf{H}$, $\mathbf{q}$, $N-k$, $e$}
  \Outputs{$\mathbf{r_c}$, $\mathbf{r_e}$, $\mathbf{H_c}$, $\mathbf{E}$}
  $\mathbf{r_e} \leftarrow \mathbf{r}[\mathbf{q}]$, ~
  $\mathbf{r_c} \leftarrow \mathbf{r} \setminus \mathbf{r_c}$\\
  $\mathbf{H_e} \leftarrow \mathbf{H}[\mathbf{q}]$, ~
  $\mathbf{H_c} \leftarrow \mathbf{H} \setminus \mathbf{H_c}$\\
  \If{$N-k < e$}{
    no unique solution, terminate decoding
  }
  $\mathbf{E} \leftarrow \text{GaussianElimination}(\mathbf{H_e}) $ \\
\end{algorithm}

\begin{algorithm}[t]
\caption{The EDGE Subroutine}\label{alg:edge}
\SetKwData{Left}{left}\SetKwData{This}{this}\SetKwData{Up}{up}
\SetKwFunction{Union}{Union}\SetKwFunction{FindCompress}{FindCompress}
\SetKwInOut{Inputs}{Inputs}\SetKwInOut{Outputs}{Outputs}
  \Inputs{$\mathbf{r_c}$, $\mathbf{r_e}$, $\mathbf{H_c}$, $\mathbf{E}$, $\mathbf{q}$, $N-k$, $e$}
  \Outputs{$\mathbf{r}$, $\textit{success}$}
  
    $\mathbf{s_e} = \mathbf{E} \cdot \mathbf{H_c} \cdot \mathbf{r_c} $ \\
    \If{$\mathbf{s_e}[e:N-k-1] \neq \mathbf{0}$}{
      \textit{success} = 0, terminate subroutine\\
    }
    $\mathbf{r_e} = \mathbf{s_e}[0:e-1]$ \\
    $\mathbf{r} \leftarrow \mathbf{r_c} \cup \mathbf{r_e}$ using $\mathbf{q}$ \\
    \textit{success} = 1\\
  
\end{algorithm}

\begin{figure*}[t]
\includegraphics[width=2.1\columnwidth]{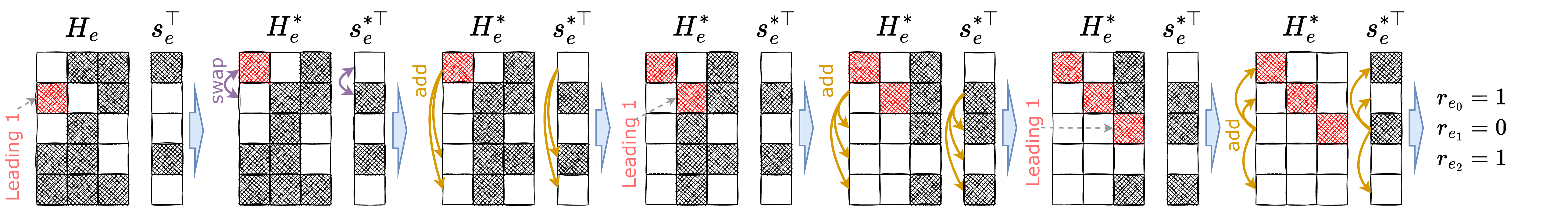}
\caption{Gaussian elimination example using two elementary row operations to transform $\mathbf{H_e}$ into reduced row echelon form (RREF), to find $\mathbf{r_e}$ from $\mathbf{s_e^*}$. 1s and 0s in the matrix and the vector are indicated by black and white, respectively. Leading 1s at each column of $\mathbf{H_e}$ are represented by red.}
\label{fig:rref}
\end{figure*}

The initialization procedure for the EDGE subroutine is described in Algorithm~\ref{alg:edge:init}, where $\mathbf{r_c}$, $\mathbf{r_e}$, $\mathbf{H_c}$ are prepared (lines 1-2) and whether the number of erasures could be recovered by the code is determined (lines 3-5). This is followed by the Gaussian elimination process using $\mathbf{H_e}$, to store the required operations in an \textit{elimination matrix}, $\mathbf{E}$ (line 6). These stored Gaussian elimination operations are later used to reduce the erasure restoration complexity, explained in Section~\ref{sec:3:ge}.

The EDGE subroutine is described in Algorithm~\ref{alg:edge}. First, the erasure syndrome is calculated $\mathbf{s_e}$ using $\mathbf{H_c}$ and $\mathbf{r_c}$, and the Gaussian elimination matrix $\mathbf{E}$. If the resulting $\mathbf{s_e}$ is \textit{error-free} (lines 2-4), then the erased sequence $\mathbf{r_e}$ is substituted from $\mathbf{s_e}$ (line 5) and the complete received sequence is restored (line 6). Otherwise, the subroutine is terminated with no success (line 3).


\subsection{The Gaussian Elimination Process}\label{sec:3:ge}

The Gaussian elimination process reduces the linear set of equations into a form from which the variables can be directly obtained. For our purpose, we review the binary matrix case of Gaussian elimination \cite{strang_book}. To achieve this form, the $\mathbf{H_e}$ matrix should be modified into a reduced row-echelon form (RREF), with the particular structure of $[\mathbf{I}|\mathbf{0}]^\top$ where $\mathbf{I}$ and $\mathbf{0}$ represent identity and all-zero matrices, respectively. Note that there is a unique RREF for any matrix. If the RREF of a matrix yields an all-zero column, there is no unique solution for $\mathbf{r_e}$.

An example of the Gaussian elimination process is depicted in Fig.~\ref{fig:rref}. Starting from the leftmost column, a leading $1$ is first identified for each column. If the row index of the leading $1$ does not match its column index, an elementary swap operation takes place to place the leading $1$ cell to the diagonal of the matrix. This row with the leading $1$ is then subtracted from the other rows that also hold a $1$ on the subject column, to ensure there is a single $1$ remaining. The process is sequentially continued for all the columns in $\mathbf{H_e}$. The equivalent swap and add operations are also applied to the transposed $\mathbf{s_e^{\top}}$, to obtain $\mathbf{s_e^{*\top}}$ at the end. Changes to the $\mathbf{H_e}$ and $\mathbf{s_e^{\top}}$ are denoted with an asterisk ($^{*}$). 

In practice, Gaussian elimination is costly with a complexity of $O(n^3)$. To reduce its impact, it can be performed only once at the initialization (refer to Algorithm~\ref{alg:edge:init}) and all the operations towards obtaining the RREF $\mathbf{H_e^*}$ can be stored in an \textit{elimination matrix}, $\mathbf{E}$. This way, the final $\mathbf{s_e^{*\top}}$ can be obtained by
\begin{equation}
    \mathbf{s_e^{*\top}} = \mathbf{E} \cdot \mathbf{s_e} \text{,}
\end{equation}
which has the same effective complexity as the syndrome check operation of GRAND, see~(\ref{eq:pc}).

\begin{figure}[t]
\includegraphics[width=1.05\columnwidth]{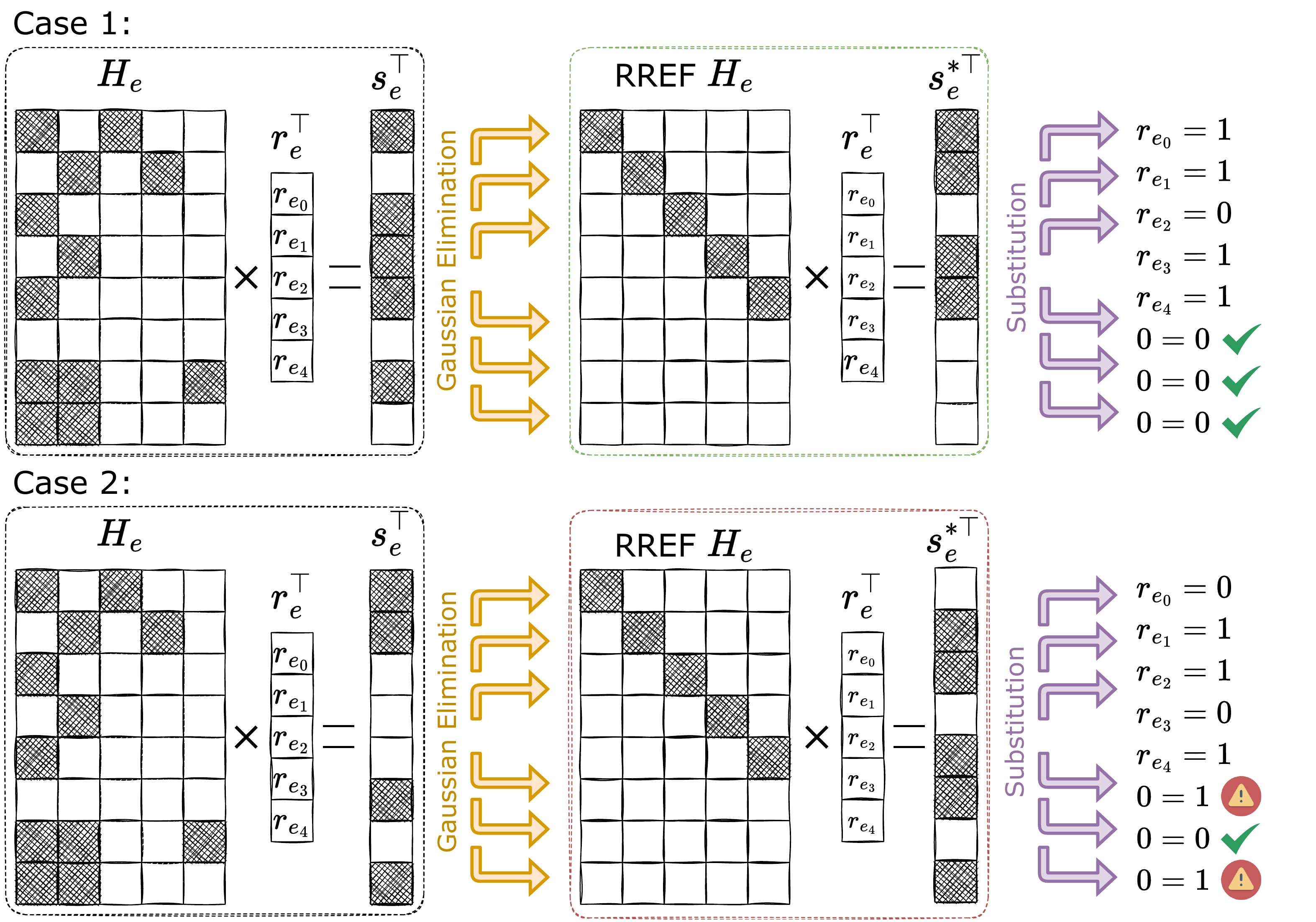}
\caption{Two examples are shown for erased bits restoration using the EDGE algorithm, with $5$ erased bits and $n-k=8$. In both cases same indices are erased from the same codebook, yielding the same $H_e$. The result of Case 1 is consistent and is therefore valid. In Case 2, the result is inconsistent due to incurred errors and therefore is invalid.}
\label{fig:example}
\end{figure}

\subsection{The GRAND-EDGE Algorithm Family}\label{sec:3:grandedge}

Refer to Fig.~\ref{fig:example}, where two arbitrary Gaussian elimination examples are illustrated using the same $\mathbf{H_e}$, with $e=5$ and $N-k=8$. In the first case, $\mathbf{s_e}$ does not incur any channel errors. Therefore, the transformed $\mathbf{s_e^*}$ is in the form of $[\mathbf{r_e} | \mathbf{0}]$, in other words, the last $N-k-e$ equations with $0$ coefficients naturally yield $0$. In the second case, the $\mathbf{s_e}$ is infused with channel errors that yield an $\mathbf{s_e^*}$ with nonzero entries in its last $N-k-e$ indices. For the EDGE subroutine, this indicates that there are errors in the channel. When errors are involved, erased indices cannot be accurately restored, and an additional error-correction decoder is required.


The process of the proposed GRAND-EDGE algorithm is described in Algorithm~\ref{alg:grand-edge}. Note that the EDGE subroutine (line 6) replaces the syndrome check function of GRAND. If the EDGE subroutine fails, then the GRAND algorithm generates the next putative error pattern and combines it with the received part of the codeword (lines 6-7, followed by line 4). As discussed in Section~\ref{sec:bg}, the agenda of the error pattern generation depends on the GRAND variant. For simplicity, inputs and parameters for soft-information GRAND variants are not shown in the algorithm, instead, we refer to \cite{solomon2020sgrand, duffy2022orbgrand} for details. 

When there are no erasures in the received codeword, then $\mathbf{r_e} = \mathbf{H_e} = \varnothing$, with $\mathbf{r_c} = \mathbf{r}$ and $\mathbf{H_c} = \mathbf{H}$. Hence, the expression in (\ref{eq:se}) becomes the same as in (\ref{eq:pc}), and GRAND-EDGE reverts to GRAND algorithm. Therefore, the proposed GRAND-EDGE does not present a computational burden to the original algorithm in the absence of erasures in the channel.

\begin{algorithm}[t]
\caption{The GRAND-EDGE Algorithm}\label{alg:grand-edge}
\SetKwData{Left}{left}\SetKwData{This}{this}\SetKwData{Up}{up}
\SetKwFunction{Union}{Union}\SetKwFunction{FindCompress}{FindCompress}
\SetKwInOut{Inputs}{Inputs}\SetKwInOut{Outputs}{Outputs}
  \Inputs{$\mathbf{r}$,$\mathbf{H}$,$\mathbf{G}^{-1}$,$\mathbf{q}$}
  \Outputs{$\mathbf{\hat{u}}$}
  $\mathbf{e} \leftarrow \mathbf{0}$ \\
  $\textit{iter} = 0$ \\
  $\{\mathbf{r_c}$, $\mathbf{r_e}$, $\mathbf{H_c}$, $\mathbf{E}\} \leftarrow$ $\text{EDGE\_Init}(\mathbf{r}, \mathbf{H}, \mathbf{q}, N-k, e)$ \\
  \While{$\textit{success} = 0 \land \textit{iter} \neq \textit{maxIters}$}{
  $\mathbf{r_c^*} \leftarrow \mathbf{r_c} \oplus \mathbf{e}$ \\
    $\{\textit{success}, \mathbf{r^*}\} \leftarrow $EDGE($\mathbf{r_c^*}$, $\mathbf{r_e}$, $\mathbf{H_c}$, $\mathbf{E}$, $\mathbf{q}$, $N-k$, $e$) \\
    $\mathbf{e} \leftarrow $NextErrorPattern$(\textit{iter})$ \\
    $\textit{iter} = \textit{iter} + 1$ \\
  }
  $\mathbf{\hat{u}} \leftarrow \mathbf{r^*} \cdot \mathbf{G}^{-1}$ \\

\end{algorithm}

\section{Performance Assessment}\label{sec:res}

Simulations are carried over the channel model described in Fig.~\ref{fig:chn}, where the channel and jammer AWGN are generated through independent Gaussian processes. The overpowered jammer SNR is set to -100 dB. The probability of jamming, $\epsilon$, is generated through an independent Bernoulli process for each transmitted symbol. Prior to decoding, a threshold is required to decide whether the received signal is jammed. For that purpose, we follow the empirical rule; any signal that is observed within $3$ standard deviations of the modulated signal space is considered non-jammed, and is jammed otherwise. Note that different approaches for determining jamming in the channel can also be used with the proposed algorithm, but are beyond the scope of this paper. For all evaluations, a random linear code (RLC) with code length $N=128$ with $k=105$ information bits is featured, which is generated using the simulator on-the-fly. The abandonment threshold \cite{duffy2019grand} is set to a Hamming weight of $3$ for both GRAND and GRAND-EDGE. The logistic weight threshold \cite{duffy2021orbgrand} is set to $104$ for ORBGRAND and ORBGRAND-EDGE.

\subsection{GRAND-EDGE Performance}\label{sec:res:grand}

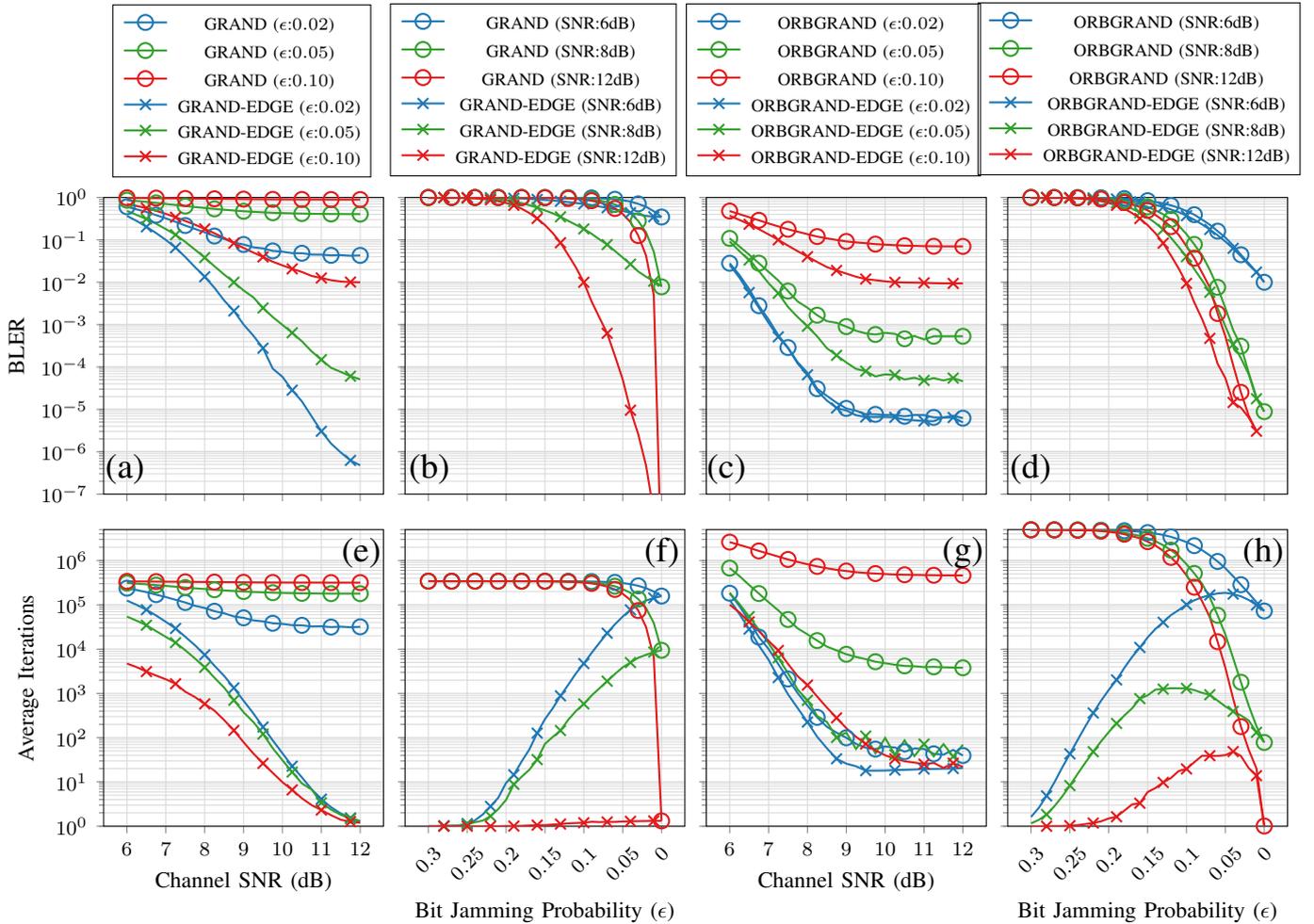
\begin{figure*}[t]
  \centering
  \input{figures/merged_grand_v2.tikz}
  \vspace{-1.25mm}
  \caption{\label{fig:res} Top: BLER performance comparison of the proposed GRAND-EDGE and ORBGRAND-EDGE algorithms against their GRAND-only counterparts, using RLC[128,105]. The comparisons are carried out with various channel AWGN SNR and bit-jamming probabilities as indicated in the x-axis labels and the legends. Bottom: Average number of iterations for each evaluated algorithm with matching legends.}
\end{figure*}

Fig.~\ref{fig:res}(a) presents the error correction performance for the proposed GRAND-EDGE algorithm against the GRAND algorithm. The x-axis represents the channel SNR, and $\epsilon$ is fixed for each simulation and is represented in the legend. We observed that GRAND-EDGE outperforms GRAND by up to five orders of magnitude, because it has the capability of restoring erased bits whereas GRAND attempts to guess their values iteratively. 

 Fig.~\ref{fig:res}(e) shows the average number of iterations (codebook queries) for each BLER curve. With improving AWGN SNR conditions, the GRAND algorithm gets stuck in guessing the jammed bits in an attempt to find the correct codeword. This yields a high average number of iterations, even at high SNRs. On the other hand, the efficient handling of the erased bits by the EDGE routine helps the GRAND-EDGE algorithm reduces the avarage number of iterations compared to GRAND, by up to more than five orders of magnitude. 

Fig.~\ref{fig:res}(b) and Fig.~\ref{fig:res}(f) present the GRAND-EDGE performance against GRAND over a simulated set of bit jamming probability values $\epsilon$. The channel SNR values are fixed and are represented in the legend. Similar to Fig.~\ref{fig:res}(a), the GRAND-EDGE algorithm achieves superior BLER performance as the channel conditions improve. On the other hand, the average number of iterations of GRAND-EDGE increases as $\epsilon$ decreases. This is because GRAND-EDGE abandons decoding if the number of erasures is beyond its capability (if $e > N-k$). Nonetheless, the average number of iterations of GRAND-EDGE is always smaller or equal to that of GRAND, sometimes by up to five orders of magnitude less. Finally, we note that both the BLER performance and the average number of iterations of GRAND-EDGE meet the performance of GRAND at $\epsilon=0$. This is because the GRAND-EDGE algorithm reverts to the GRAND algorithm when there is no erasure in the channel, as mentioned previously in Section~\ref{sec:3:grandedge}.


\subsection{ORBGRAND-EDGE Performance}\label{sec:res:orbgrand}

Fig.~\ref{fig:res}(c) and Fig.~\ref{fig:res}(d) present the BLER performance evaluation for the proposed ORBGRAND-EDGE algorithm against the ORBGRAND algorithm, \textit{i.e.} when soft-information is involved in decoding the received codeword. Compared to GRAND, the ORBGRAND algorithm can find more errors by taking advantage of the bit-reliability information (LLRs). As a result, ORBGRAND demonstrates better BLER performance compared to GRAND. On the other hand, the improvement in BLER with ORBGRAND-EDGE is not as dramatic compared to the gains in hard-information GRAND-EDGE. This is because ORBGRAND can prioritize jammed indices for bit-flipping as long as their values are beyond the predetermined threshold. Nonetheless, we observe a BLER gain of up to one order of magnitude when ORBGRAND is enhanced with the EDGE subroutine. 

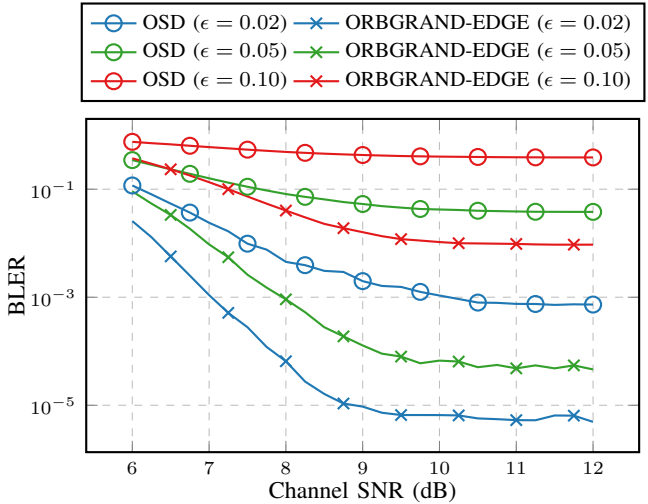
\begin{figure}[t]
  \centering
  \input{figures/results_osd_p_bler.tikz}
   \caption{\label{fig:res_osd_fer} ORBGRAND-EDGE performance compared to the OSD algorithm, using RLC[128,105]. The bit-level jamming probabilities are set to $\epsilon = \{0.02, 0.05, 0.1\}$ and x-axis represents the channel AWGN SNR.}
\end{figure}

Compared to the GRAND-EDGE performance in Fig.~\ref{fig:res}(a) with $\epsilon = 0.02$, the ORBGRAND-EDGE performance in Fig.~\ref{fig:res}(c) experiences an error floor. As a result, at high SNR regimes, GRAND-EDGE outperforms ORBGRAND-EDGE. This is due to the differences in the bit-flipping agenda of these two variants. With a logistic weight of $104$, only the least reliable $104$ indices and a subset of their combinations are considered for bit-flipping in ORBGRAND \cite{duffy2022orbgrand}. Therefore, when a jammed index is overlooked by the jamming detection, ORBGRAND-EDGE does not erase the bit, and it is likely considered an index reliable enough that it is never considered for bit-flipping. On the other hand, even when a jammed index is overseen, the GRAND-EDGE algorithm evaluates all non-erased indices for bit-flipping. As a result, especially when the $\epsilon$ is small, GRAND-EDGE can outperform ORBGRAND-EDGE. Although more sophisticated pre-decoding jamming detection patterns can be implemented \cite{ercan_jamming}, they are beyond the scope of this work.

Fig.~\ref{fig:res}(g) and Fig.~\ref{fig:res}(h) present the computational complexity comparison for ORBGRAND-EDGE and ORBGRAND, with respect to channel SNR and bit-jamming probability, respectively. Compared to GRAND, ORBGRAND has better (less) computational complexity in general, due to the efficient identification of bit-flipping indices. On the other hand, when ORBGRAND is augmented with the EDGE subroutine, the average number of iterations reduces by up to five orders of magnitude. Similar to the observations made in GRAND, ORBGRAND-EDGE reverts to the ORBGRAND algorithm when $\epsilon=0$. Different than GRAND-EDGE, the computational complexity of ORBGRAND-EDGE begins to reduce as $\epsilon$ reduces further. This is because the complexity of the linear set of equations reduces with reducing $\epsilon$, and with the help of soft information ORBGRAND-EDGE can find the erroneous component of the received codeword faster compared to GRAND-EDGE.

Fig.~\ref{fig:res_osd_fer} compares the BLER performance of the proposed ORBGRAND-EDGE algorithm against the Ordered Statistics Decoding (OSD) algorithm \cite{osd1995}, using RLC[128,105]. Similar to EDGE, OSD performs Gaussian elimination to find the most likely transmitted codeword. However, the Gaussian elimination in OSD is always performed over $k$ columns, whereas in EDGE it is only performed over up to $N-k$ columns. Moreover, OSD requires multiple permutations which adds to its implementation complexity. Nonetheless, it can be observed from Fig.~\ref{fig:res_osd_fer} that the ORBGRAND-EDGE outperforms OSD by up to three orders of magnitude in BLER.

\section{Conclusion}\label{sec:conc}

In this work, we introduced an adversarial model, whereby a jammer randomly overpowers bits of a transmitted signal, effectively causing erasures. To address this adversarial channel condition, we generalized the syndrome check component of the universal GRAND algorithm family to support erasure decoding. The proposed GRAND-EDGE algorithm and its variants address bit-level errors and erasures simultaneously. The erasure decoding component (i.e., the EDGE subroutine) presents no additional computational complexity when there is no detected erasure in the channel, reverting to the syndrome check function of the GRAND algorithm in that case. The proposed EDGE algorithm can be used with both hard and soft variants of GRAND, which we demonstrated through the implementation of GRAND-EDGE and ORBGRAND-EDGE. Compared to their original counterparts, the EDGE-enhanced GRAND algorithms achieve up to five orders of magnitude improvement both in terms of error-correction performance in terms of computational complexity under the considered adversarial model.

\section*{Acknowledgements}
This work was partially supported by Defense Advanced
Research Projects Agency Contract number HR00112120008 and by National Science Foundation ECCS Award numbers 2128517 and 2128555. The content of the information does not necessarily reflect the position or the policy of the US Government, and no official endorsement should be inferred. This publication has emanated from research supported in part by a grant from Science Foundation Ireland under grant number 18/CRT/6049. The opinions, findings and conclusions or recommendations expressed in this material are those of the author(s) and do not necessarily reflect the views of the Science Foundation Ireland.

\bibliographystyle{IEEEtran}
\bibliography{IEEEabrv,ref}

\end{document}

%% file: colors.tex
\definecolor{ocre}{RGB}{0, 157, 224} 
\definecolor{Paired-2}{RGB}{166,206,227}
\definecolor{Paired-1}{RGB}{31,120,180}
\definecolor{Paired-4}{RGB}{178,223,138}
\definecolor{Paired-3}{RGB}{51,160,44}
\definecolor{Paired-6}{RGB}{251,154,153}
\definecolor{Paired-5}{RGB}{227,26,28}
\definecolor{Paired-8}{RGB}{253,191,111}
\definecolor{Paired-7}{RGB}{255,127,0}
\definecolor{Paired-10}{RGB}{202,178,214}
\definecolor{Paired-9}{RGB}{106,61,154}
\definecolor{Paired-12}{RGB}{255,255,153}
\definecolor{Paired-11}{RGB}{177,89,40}
\definecolor{Accent-1}{RGB}{127,201,127}
\definecolor{Accent-2}{RGB}{190,174,212}
\definecolor{Accent-3}{RGB}{253,192,134}
\definecolor{Accent-4}{RGB}{255,255,153}
\definecolor{Accent-5}{RGB}{56,108,176}
\definecolor{Accent-6}{RGB}{240,2,127}
\definecolor{Accent-7}{RGB}{191,91,23}
\definecolor{Accent-8}{RGB}{102,102,102}
\definecolor{Spectral-1}{RGB}{158,1,66}
\definecolor{Spectral-2}{RGB}{213,62,79}
\definecolor{Spectral-3}{RGB}{244,109,67}
\definecolor{Spectral-4}{RGB}{253,174,97}
\definecolor{Spectral-5}{RGB}{254,224,139}
\definecolor{Spectral-6}{RGB}{255,255,191}
\definecolor{Spectral-7}{RGB}{230,245,152}
\definecolor{Spectral-8}{RGB}{171,221,164}
\definecolor{Spectral-9}{RGB}{102,194,165}
\definecolor{Spectral-10}{RGB}{50,136,189}
\definecolor{Spectral-11}{RGB}{94,79,162}
\definecolor{Set1-1}{RGB}{228,26,28}
\definecolor{Set1-2}{RGB}{55,126,184}
\definecolor{Set1-3}{RGB}{77,175,74}
\definecolor{Set1-4}{RGB}{152,78,163}
\definecolor{Set1-5}{RGB}{255,127,0}
\definecolor{Set1-6}{RGB}{255,255,51}
\definecolor{Set1-7}{RGB}{166,86,40}
\definecolor{Set1-8}{RGB}{247,129,191}
\definecolor{Set1-9}{RGB}{153,153,153}
\definecolor{Set2-1}{RGB}{102,194,165}
\definecolor{Set2-2}{RGB}{252,141,98}
\definecolor{Set2-3}{RGB}{141,160,203}
\definecolor{Set2-4}{RGB}{231,138,195}
\definecolor{Set2-5}{RGB}{166,216,84}
\definecolor{Set2-6}{RGB}{255,217,47}
\definecolor{Set2-7}{RGB}{229,196,148}
\definecolor{Set2-8}{RGB}{179,179,179}
\definecolor{Dark2-1}{RGB}{27,158,119}
\definecolor{Dark2-2}{RGB}{217,95,2}
\definecolor{Dark2-3}{RGB}{117,112,179}
\definecolor{Dark2-4}{RGB}{231,41,138}
\definecolor{Dark2-5}{RGB}{102,166,30}
\definecolor{Dark2-6}{RGB}{230,171,2}
\definecolor{Dark2-7}{RGB}{166,118,29}
\definecolor{Dark2-8}{RGB}{102,102,102}
\definecolor{Reds-1}{RGB}{255,245,240}
\definecolor{Reds-2}{RGB}{254,224,210}
\definecolor{Reds-3}{RGB}{252,187,161}
\definecolor{Reds-4}{RGB}{252,146,114}
\definecolor{Reds-5}{RGB}{251,106,74}
\definecolor{Reds-6}{RGB}{239,59,44}
\definecolor{Reds-7}{RGB}{203,24,29}
\definecolor{Reds-8}{RGB}{165,15,21}
\definecolor{Reds-9}{RGB}{103,0,13}
\definecolor{Greens-1}{RGB}{247,252,245}
\definecolor{Greens-2}{RGB}{229,245,224}
\definecolor{Greens-3}{RGB}{199,233,192}
\definecolor{Greens-4}{RGB}{161,217,155}
\definecolor{Greens-5}{RGB}{116,196,118}
\definecolor{Greens-6}{RGB}{65,171,93}
\definecolor{Greens-7}{RGB}{35,139,69}
\definecolor{Greens-8}{RGB}{0,109,44}
\definecolor{Greens-9}{RGB}{0,68,27}
\definecolor{Blues-1}{RGB}{247,251,255}
\definecolor{Blues-2}{RGB}{222,235,247}
\definecolor{Blues-3}{RGB}{198,219,239}
\definecolor{Blues-4}{RGB}{158,202,225}
\definecolor{Blues-5}{RGB}{107,174,214}
\definecolor{Blues-6}{RGB}{66,146,198}
\definecolor{Blues-7}{RGB}{33,113,181}
\definecolor{Blues-8}{RGB}{8,81,156}
\definecolor{Blues-9}{RGB}{8,48,107}
\definecolor{awesome-skyblue}{HTML}{0395DE}
\definecolor{bg}{HTML}{282828}

\definecolor{rosso}{RGB}{220,57,18}
\definecolor{giallo}{RGB}{255,153,0}
\definecolor{blu}{RGB}{102,140,217}
\definecolor{verde}{RGB}{16,150,24}
\definecolor{viola}{RGB}{153,0,153}

\definecolor{bleuUni}{RGB}{0, 157, 224}
\definecolor{marronUni}{RGB}{68, 58, 49}

%% file: figures/merged_grand_v2.tikz
\begin{tikzpicture}

    \begin{groupplot}[
        group style={group name=group, group size= 4 by 2, horizontal sep=.3cm, vertical sep=2.0cm},
        footnotesize, width=2*\columnwidth, height=0.65\columnwidth,
        ymode=log,
        ymin = 1.0e-07, ymax=1,
        grid=both, grid style={gray!30},
        tick align=outside, tickpos=left, 
        ]

\nextgroupplot[ylabel={BLER}, xticklabels={,,}, width = .62\columnwidth]

\addplot[
    color=Paired-1,
    mark=o,
    thick,
    mark size=3,
    mark repeat=3,mark phase=1,
]
table {
6.000e+00 6.04100e-01
6.250e+00 5.25100e-01
6.500e+00 4.55400e-01
6.750e+00 3.83200e-01
7.000e+00 3.21600e-01
7.250e+00 2.66900e-01
7.500e+00 2.17900e-01
7.750e+00 1.80900e-01
8.000e+00 1.50800e-01
8.250e+00 1.23100e-01
8.500e+00 1.05500e-01
8.750e+00 8.63000e-02
9.000e+00 7.79000e-02
9.250e+00 6.68000e-02
9.500e+00 6.29000e-02
9.750e+00 5.49000e-02
1.000e+01 5.27000e-02
1.025e+01 4.87000e-02
1.050e+01 4.84000e-02
1.075e+01 4.50000e-02
1.100e+01 4.48000e-02
1.125e+01 4.32000e-02
1.150e+01 4.37000e-02
1.175e+01 4.23000e-02
1.200e+01 4.30000e-02
};
\label{gp:g_e2}

\addplot[
    color=Paired-1,
    mark=x,
    thick,
    mark size=3,
    mark repeat=3,mark phase=3,
]
table {
6.000e+00 3.67800e-01
6.250e+00 2.79500e-01
6.500e+00 2.03400e-01
6.750e+00 1.44800e-01
7.000e+00 9.96000e-02
7.250e+00 6.55000e-02
7.500e+00 3.94000e-02
7.750e+00 2.27000e-02
8.000e+00 1.35000e-02
8.250e+00 7.40000e-03
8.500e+00 3.71581e-03
8.750e+00 2.10181e-03
9.000e+00 1.00539e-03
9.250e+00 5.49626e-04
9.500e+00 2.77994e-04
9.750e+00 9.15091e-05
1.000e+01 5.86428e-05
1.025e+01 2.83896e-05
1.050e+01 1.54321e-05
1.075e+01 6.82998e-06
1.100e+01 3.03745e-06
1.125e+01 1.56048e-06
1.150e+01 9.60606e-07
1.175e+01 6.23711e-07
1.200e+01 4.78074e-07
};
\label{gp:ge_e2}

\addplot[
    color=Paired-3,
    mark=o,
    thick,
    mark size=3,
    mark repeat=3,mark phase=1,
]
table {
6.000e+00 8.63200e-01
6.250e+00 8.26500e-01
6.500e+00 7.88100e-01
6.750e+00 7.45700e-01
7.000e+00 7.03600e-01
7.250e+00 6.64700e-01
7.500e+00 6.27100e-01
7.750e+00 5.95100e-01
8.000e+00 5.65500e-01
8.250e+00 5.38800e-01
8.500e+00 5.14200e-01
8.750e+00 4.92800e-01
9.000e+00 4.73900e-01
9.250e+00 4.59100e-01
9.500e+00 4.46300e-01
9.750e+00 4.36600e-01
1.000e+01 4.28800e-01
1.025e+01 4.22500e-01
1.050e+01 4.17900e-01
1.075e+01 4.12400e-01
1.100e+01 4.10000e-01
1.125e+01 4.07600e-01
1.150e+01 4.06600e-01
1.175e+01 4.05600e-01
1.200e+01 4.05100e-01
};
\label{gp:g_e5}

\addplot[
    color=Paired-3,
    mark=x,
    thick,
    mark size=3,
    mark repeat=3,mark phase=3,
]
table {
6.000e+00 4.81500e-01
6.250e+00 3.88000e-01
6.500e+00 3.10700e-01
6.750e+00 2.40800e-01
7.000e+00 1.79600e-01
7.250e+00 1.32000e-01
7.500e+00 8.96000e-02
7.750e+00 5.88000e-02
8.000e+00 3.82000e-02
8.250e+00 2.38000e-02
8.500e+00 1.52000e-02
8.750e+00 1.00000e-02
9.000e+00 6.40000e-03
9.250e+00 4.41151e-03
9.500e+00 2.47194e-03
9.750e+00 1.51021e-03
1.000e+01 9.83845e-04
1.025e+01 6.41626e-04
1.050e+01 4.03916e-04
1.075e+01 2.33598e-04
1.100e+01 1.48899e-04
1.125e+01 9.57170e-05
1.150e+01 7.65703e-05
1.175e+01 6.06516e-05
1.200e+01 5.11339e-05
};
\label{gp:ge_e5}

\addplot[
    color=Paired-5,
    mark=o,
    thick,
    mark size=3,
    mark repeat=3,mark phase=1,
]
table {
6.000e+00 9.85500e-01
6.250e+00 9.79500e-01
6.500e+00 9.74200e-01
6.750e+00 9.67700e-01
7.000e+00 9.60800e-01
7.250e+00 9.54700e-01
7.500e+00 9.48400e-01
7.750e+00 9.43500e-01
8.000e+00 9.38100e-01
8.250e+00 9.31500e-01
8.500e+00 9.26300e-01
8.750e+00 9.21200e-01
9.000e+00 9.15900e-01
9.250e+00 9.10900e-01
9.500e+00 9.06800e-01
9.750e+00 9.03900e-01
1.000e+01 9.02600e-01
1.025e+01 9.00400e-01
1.050e+01 8.98800e-01
1.075e+01 8.97500e-01
1.100e+01 8.96500e-01
1.125e+01 8.95700e-01
1.150e+01 8.95200e-01
1.175e+01 8.95200e-01
1.200e+01 8.95000e-01
};
\label{gp:g_e10}

\addplot[
    color=Paired-5,
    mark=x,
    thick,
    mark size=3,
    mark repeat=3,mark phase=3,
]
table {
6.000e+00 7.02900e-01
6.250e+00 6.31800e-01
6.500e+00 5.60000e-01
6.750e+00 4.84100e-01
7.000e+00 4.11400e-01
7.250e+00 3.41900e-01
7.500e+00 2.79500e-01
7.750e+00 2.28000e-01
8.000e+00 1.81400e-01
8.250e+00 1.43500e-01
8.500e+00 1.11000e-01
8.750e+00 8.36000e-02
9.000e+00 6.45000e-02
9.250e+00 5.02000e-02
9.500e+00 3.96000e-02
9.750e+00 3.05000e-02
1.000e+01 2.50000e-02
1.025e+01 2.07000e-02
1.050e+01 1.75000e-02
1.075e+01 1.41000e-02
1.100e+01 1.26000e-02
1.125e+01 1.12000e-02
1.150e+01 1.04000e-02
1.175e+01 1.01000e-02
1.200e+01 1.00000e-02
};
\label{gp:ge_e10}

\coordinate (top) at (rel axis cs:0,0);

\nextgroupplot[yticklabels={,,}, xticklabels={,,}, width = .62\columnwidth,  xtick={0.3, 0.25, 0.2, 0.15, 0.10, 0.05, 0},x dir=reverse]

\addplot[
    color=Paired-1,
    mark=o,
    thick,
    mark size=3,
    mark repeat=3,mark phase=1,
]
table {
3.000e-01 1.00000e+00
2.900e-01 1.00000e+00
2.800e-01 1.00000e+00
2.700e-01 1.00000e+00
2.600e-01 1.00000e+00
2.500e-01 1.00000e+00
2.400e-01 1.00000e+00
2.300e-01 1.00000e+00
2.200e-01 1.00000e+00
2.100e-01 9.99900e-01
2.000e-01 1.00000e+00
1.900e-01 9.99800e-01
1.800e-01 9.99700e-01
1.700e-01 9.99700e-01
1.600e-01 9.99100e-01
1.500e-01 9.98900e-01
1.400e-01 9.97700e-01
1.300e-01 9.95500e-01
1.200e-01 9.93700e-01
1.100e-01 9.91000e-01
1.000e-01 9.85500e-01
9.000e-02 9.75700e-01
8.000e-02 9.60200e-01
7.000e-02 9.35400e-01
6.000e-02 9.07000e-01
5.000e-02 8.60200e-01
4.000e-02 8.02700e-01
3.000e-02 7.07500e-01
2.000e-02 6.04100e-01
1.000e-02 4.73300e-01
0.000e+00 3.46500e-01
};
\label{gp:g_s6}

\addplot[
    color=Paired-1,
    mark=x,
    thick,
    mark size=3,
    mark repeat=3,mark phase=3,
]
table {
3.000e-01 9.99900e-01
2.900e-01 9.99800e-01
2.800e-01 9.99800e-01
2.700e-01 9.99600e-01
2.600e-01 9.98600e-01
2.500e-01 9.97400e-01
2.400e-01 9.95000e-01
2.300e-01 9.92500e-01
2.200e-01 9.88100e-01
2.100e-01 9.81500e-01
2.000e-01 9.73600e-01
1.900e-01 9.60100e-01
1.800e-01 9.45600e-01
1.700e-01 9.25400e-01
1.600e-01 9.02900e-01
1.500e-01 8.78900e-01
1.400e-01 8.51000e-01
1.300e-01 8.15800e-01
1.200e-01 7.79700e-01
1.100e-01 7.41400e-01
1.000e-01 7.02900e-01
9.000e-02 6.57100e-01
8.000e-02 6.12700e-01
7.000e-02 5.71500e-01
6.000e-02 5.26000e-01
5.000e-02 4.81500e-01
4.000e-02 4.35500e-01
3.000e-02 3.98300e-01
2.000e-02 3.67800e-01
1.000e-02 3.51700e-01
0.000e+00 3.46900e-01
};
\label{gp:ge_s6}

\addplot[
    color=Paired-3,
    mark=o,
    thick,
    mark size=3,
    mark repeat=3,mark phase=1,
]
table {
3.000e-01 1.00000e+00
2.900e-01 1.00000e+00
2.800e-01 1.00000e+00
2.700e-01 1.00000e+00
2.600e-01 1.00000e+00
2.500e-01 1.00000e+00
2.400e-01 9.99900e-01
2.300e-01 1.00000e+00
2.200e-01 1.00000e+00
2.100e-01 9.99700e-01
2.000e-01 9.99700e-01
1.900e-01 9.99400e-01
1.800e-01 9.99200e-01
1.700e-01 9.97800e-01
1.600e-01 9.96300e-01
1.500e-01 9.94700e-01
1.400e-01 9.89400e-01
1.300e-01 9.81900e-01
1.200e-01 9.69400e-01
1.100e-01 9.58300e-01
1.000e-01 9.38100e-01
9.000e-02 9.00700e-01
8.000e-02 8.48300e-01
7.000e-02 7.76500e-01
6.000e-02 6.85400e-01
5.000e-02 5.61800e-01
4.000e-02 4.30800e-01
3.000e-02 2.79700e-01
2.000e-02 1.50800e-01
1.000e-02 5.24000e-02
0.000e+00 7.80000e-03
};
\label{gp:g_s8}

\addplot[
    color=Paired-3,
    mark=x,
    thick,
    mark size=3,
    mark repeat=3,mark phase=3,
]
table {
3.000e-01 9.99600e-01
2.900e-01 9.99300e-01
2.800e-01 9.98500e-01
2.700e-01 9.96900e-01
2.600e-01 9.93700e-01
2.500e-01 9.86300e-01
2.400e-01 9.74300e-01
2.300e-01 9.55400e-01
2.200e-01 9.33300e-01
2.100e-01 8.98900e-01
2.000e-01 8.57700e-01
1.900e-01 8.03400e-01
1.800e-01 7.35900e-01
1.700e-01 6.56100e-01
1.600e-01 5.76300e-01
1.500e-01 4.96600e-01
1.400e-01 4.23400e-01
1.300e-01 3.49700e-01
1.200e-01 2.80300e-01
1.100e-01 2.25900e-01
1.000e-01 1.81400e-01
9.000e-02 1.39300e-01
8.000e-02 1.03100e-01
7.000e-02 7.74000e-02
6.000e-02 5.47000e-02
5.000e-02 3.82000e-02
4.000e-02 2.67000e-02
3.000e-02 1.79000e-02
2.000e-02 1.35000e-02
1.000e-02 1.06000e-02
0.000e+00 8.20000e-03
};
\label{gp:ge_s8}

\addplot[
    color=Paired-5,
    mark=o,
    thick,
    mark size=3,
    mark repeat=3,mark phase=1,
]
table {
3.000e-01 1.00000e+00
2.900e-01 1.00000e+00
2.800e-01 1.00000e+00
2.700e-01 9.99900e-01
2.600e-01 1.00000e+00
2.500e-01 1.00000e+00
2.400e-01 9.99900e-01
2.300e-01 9.99900e-01
2.200e-01 9.99900e-01
2.100e-01 9.99100e-01
2.000e-01 9.99300e-01
1.900e-01 9.98400e-01
1.800e-01 9.97800e-01
1.700e-01 9.95900e-01
1.600e-01 9.93100e-01
1.500e-01 9.89500e-01
1.400e-01 9.81600e-01
1.300e-01 9.69400e-01
1.200e-01 9.50700e-01
1.100e-01 9.31700e-01
1.000e-01 8.95000e-01
9.000e-02 8.39700e-01
8.000e-02 7.68500e-01
7.000e-02 6.69800e-01
6.000e-02 5.50400e-01
5.000e-02 4.09300e-01
4.000e-02 2.62700e-01
3.000e-02 1.26200e-01
2.000e-02 4.30000e-02
1.000e-02 5.00000e-03
0.000e+00 1.52137e-09
};
\label{gp:g_s12}

\addplot[
    color=Paired-5,
    mark=x,
    thick,
    mark size=3,
    mark repeat=3,mark phase=3,
]
table {
3.000e-01 9.98900e-01
2.900e-01 9.97800e-01
2.800e-01 9.96000e-01
2.700e-01 9.92800e-01
2.600e-01 9.87000e-01
2.500e-01 9.73000e-01
2.400e-01 9.52300e-01
2.300e-01 9.21600e-01
2.200e-01 8.79600e-01
2.100e-01 8.19100e-01
2.000e-01 7.43700e-01
1.900e-01 6.54300e-01
1.800e-01 5.45100e-01
1.700e-01 4.29400e-01
1.600e-01 3.19700e-01
1.500e-01 2.24200e-01
1.400e-01 1.44900e-01
1.300e-01 8.55000e-02
1.200e-01 4.40000e-02
1.100e-01 2.21000e-02
1.000e-01 1.00000e-02
9.000e-02 3.52162e-03
8.000e-02 1.47427e-03
7.000e-02 6.23947e-04
6.000e-02 1.83759e-04
5.000e-02 5.11339e-05
4.000e-02 9.53948e-06
3.000e-02 2.58443e-06
2.000e-02 4.78074e-07
1.000e-02 3.68074e-08
0.000e+00 1.52137e-09
};
\label{gp:ge_s12}

\coordinate (bot) at (rel axis cs:1,0);

\nextgroupplot[yticklabels={,,}, xticklabels={,,}, width = .62\columnwidth, xtick={6, 7, ..., 12},]
\addplot[
    color=Paired-1,
    mark=o,
    thick,
    mark size=3,
    mark repeat=3,mark phase=1,
]
table {
6.000e+00 2.81000e-02
6.250e+00 1.48000e-02
6.500e+00 6.90000e-03
6.750e+00 2.79893e-03
7.000e+00 1.32300e-03
7.250e+00 5.62012e-04
7.500e+00 2.87565e-04
7.750e+00 1.28742e-04
8.000e+00 7.30028e-05
8.250e+00 3.07177e-05
8.500e+00 2.08875e-05
8.750e+00 1.43143e-05
9.000e+00 1.05212e-05
9.250e+00 9.28156e-06
9.500e+00 7.69948e-06
9.750e+00 7.62787e-06
1.000e+01 7.34556e-06
1.025e+01 7.31979e-06
1.050e+01 6.86061e-06
1.075e+01 7.34556e-06
1.100e+01 7.31979e-06
1.125e+01 6.46918e-06
1.150e+01 6.39067e-06
1.175e+01 7.11382e-06
1.200e+01 6.19090e-06
};

\addplot[
    color=Paired-1,
    mark=x,
    thick,
    mark size=3,
    mark repeat=3,mark phase=3,
]
table {
6.000e+00 2.57000e-02
6.250e+00 1.29000e-02
6.500e+00 5.70000e-03
6.750e+00 2.50803e-03
7.000e+00 1.09572e-03
7.250e+00 5.13780e-04
7.500e+00 2.78413e-04
7.750e+00 1.19863e-04
8.000e+00 6.52846e-05
8.250e+00 2.74445e-05
8.500e+00 1.61963e-05
8.750e+00 1.07768e-05
9.000e+00 9.43975e-06
9.250e+00 7.27315e-06
9.500e+00 6.59435e-06
9.750e+00 6.59435e-06
1.000e+01 6.57762e-06
1.025e+01 6.48311e-06
1.050e+01 5.70818e-06
1.075e+01 5.52828e-06
1.100e+01 5.30363e-06
1.125e+01 5.27909e-06
1.150e+01 6.46918e-06
1.175e+01 6.42566e-06
1.200e+01 4.92720e-06
};

\addplot[
    color=Paired-3,
    mark=o,
    thick,
    mark size=3,
    mark repeat=3,mark phase=1,
]
table {
6.000e+00 1.08400e-01
6.250e+00 7.00000e-02
6.500e+00 4.49000e-02
6.750e+00 2.82000e-02
7.000e+00 1.76000e-02
7.250e+00 1.10000e-02
7.500e+00 6.20000e-03
7.750e+00 3.50705e-03
8.000e+00 2.50803e-03
8.250e+00 1.66934e-03
8.500e+00 1.20674e-03
8.750e+00 1.10239e-03
9.000e+00 8.98376e-04
9.250e+00 7.16148e-04
9.500e+00 6.21504e-04
9.750e+00 5.85569e-04
1.000e+01 6.39623e-04
1.025e+01 6.07629e-04
1.050e+01 4.62937e-04
1.075e+01 5.41594e-04
1.100e+01 4.45935e-04
1.125e+01 5.30667e-04
1.150e+01 5.30667e-04
1.175e+01 5.30667e-04
1.200e+01 5.30667e-04
};

\addplot[
    color=Paired-3,
    mark=x,
    thick,
    mark size=3,
    mark repeat=3,mark phase=3,
]
table {
6.000e+00 9.13000e-02
6.250e+00 5.41000e-02
6.500e+00 3.34000e-02
6.750e+00 1.85000e-02
7.000e+00 9.50000e-03
7.250e+00 5.50000e-03
7.500e+00 2.60974e-03
7.750e+00 1.50376e-03
8.000e+00 9.13977e-04
8.250e+00 5.28290e-04
8.500e+00 2.78152e-04
8.750e+00 1.88534e-04
9.000e+00 1.28064e-04
9.250e+00 9.02122e-05
9.500e+00 7.96213e-05
9.750e+00 5.99304e-05
1.000e+01 6.70632e-05
1.025e+01 6.44366e-05
1.050e+01 5.06457e-05
1.075e+01 5.57468e-05
1.100e+01 4.82110e-05
1.125e+01 5.47406e-05
1.150e+01 4.82110e-05
1.175e+01 5.47406e-05
1.200e+01 4.60444e-05
};

\addplot[
    color=Paired-5,
    mark=o,
    thick,
    mark size=3,
    mark repeat=3,mark phase=1,
]
table {
6.000e+00 4.78800e-01
6.250e+00 4.05100e-01
6.500e+00 3.43300e-01
6.750e+00 2.89600e-01
7.000e+00 2.44300e-01
7.250e+00 2.07800e-01
7.500e+00 1.77800e-01
7.750e+00 1.55800e-01
8.000e+00 1.36400e-01
8.250e+00 1.19700e-01
8.500e+00 1.08400e-01
8.750e+00 9.97000e-02
9.000e+00 9.18000e-02
9.250e+00 8.73000e-02
9.500e+00 8.34000e-02
9.750e+00 7.94000e-02
1.000e+01 7.72000e-02
1.025e+01 7.50000e-02
1.050e+01 7.30000e-02
1.075e+01 7.22000e-02
1.100e+01 7.11000e-02
1.125e+01 7.07000e-02
1.150e+01 7.03000e-02
1.175e+01 7.03000e-02
1.200e+01 7.02000e-02
};

\addplot[
    color=Paired-5,
    mark=x,
    thick,
    mark size=3,
    mark repeat=3,mark phase=3,
]
table {
6.000e+00 3.72900e-01
6.250e+00 2.95600e-01
6.500e+00 2.33500e-01
6.750e+00 1.79300e-01
7.000e+00 1.34800e-01
7.250e+00 1.00100e-01
7.500e+00 7.32000e-02
7.750e+00 5.39000e-02
8.000e+00 4.02000e-02
8.250e+00 3.01000e-02
8.500e+00 2.28000e-02
8.750e+00 1.90000e-02
9.000e+00 1.60000e-02
9.250e+00 1.35000e-02
9.500e+00 1.19000e-02
9.750e+00 1.12000e-02
1.000e+01 1.05000e-02
1.025e+01 1.00000e-02
1.050e+01 9.90000e-03
1.075e+01 9.80000e-03
1.100e+01 9.70000e-03
1.125e+01 9.50000e-03
1.150e+01 9.40000e-03
1.175e+01 9.40000e-03
1.200e+01 9.40000e-03
};

\coordinate (top) at (rel axis cs:0,0);

\nextgroupplot[yticklabels={,,}, xticklabels={,,}, width = .62\columnwidth,  xtick={0.3, 0.25, 0.2, 0.15, 0.10, 0.05, 0}, x dir=reverse]

\addplot[
    color=Paired-1,
    mark=o,
    thick,
    mark size=3,
    mark repeat=3,mark phase=1,
]
table {
3.000e-01 1.00000e+00
2.900e-01 9.99900e-01
2.800e-01 9.99600e-01
2.700e-01 9.99400e-01
2.600e-01 9.98800e-01
2.500e-01 9.97700e-01
2.400e-01 9.97500e-01
2.300e-01 9.94800e-01
2.200e-01 9.90400e-01
2.100e-01 9.84300e-01
2.000e-01 9.76000e-01
1.900e-01 9.67600e-01
1.800e-01 9.44700e-01
1.700e-01 9.20900e-01
1.600e-01 8.87200e-01
1.500e-01 8.41400e-01
1.400e-01 7.85100e-01
1.300e-01 7.26600e-01
1.200e-01 6.51400e-01
1.100e-01 5.72000e-01
1.000e-01 4.78800e-01
9.000e-02 3.93100e-01
8.000e-02 3.00800e-01
7.000e-02 2.27400e-01
6.000e-02 1.60700e-01
5.000e-02 1.13300e-01
4.000e-02 7.37000e-02
3.000e-02 4.46000e-02
2.000e-02 2.81000e-02
1.000e-02 1.74000e-02
0.000e+00 1.00000e-02
};

\addplot[
    color=Paired-1,
    mark=x,
    thick,
    mark size=3,
    mark repeat=3,mark phase=3,
]
table {
3.000e-01 9.99900e-01
2.900e-01 9.99800e-01
2.800e-01 9.99800e-01
2.700e-01 9.99400e-01
2.600e-01 9.97900e-01
2.500e-01 9.96200e-01
2.400e-01 9.92900e-01
2.300e-01 9.88100e-01
2.200e-01 9.79900e-01
2.100e-01 9.66300e-01
2.000e-01 9.48200e-01
1.900e-01 9.20300e-01
1.800e-01 8.86100e-01
1.700e-01 8.40600e-01
1.600e-01 7.92700e-01
1.500e-01 7.32200e-01
1.400e-01 6.68500e-01
1.300e-01 5.97600e-01
1.200e-01 5.21600e-01
1.100e-01 4.49400e-01
1.000e-01 3.72900e-01
9.000e-02 2.97800e-01
8.000e-02 2.31800e-01
7.000e-02 1.76200e-01
6.000e-02 1.28700e-01
5.000e-02 9.13000e-02
4.000e-02 6.32000e-02
3.000e-02 4.09000e-02
2.000e-02 2.58000e-02
1.000e-02 1.74000e-02
0.000e+00 1.00000e-02
};

\addplot[
    color=Paired-3,
    mark=o,
    thick,
    mark size=3,
    mark repeat=3,mark phase=1,
]
table {
3.000e-01 9.99600e-01
2.900e-01 9.99600e-01
2.800e-01 9.98600e-01
2.700e-01 9.98000e-01
2.600e-01 9.95400e-01
2.500e-01 9.90300e-01
2.400e-01 9.86800e-01
2.300e-01 9.79700e-01
2.200e-01 9.64600e-01
2.100e-01 9.44900e-01
2.000e-01 9.15900e-01
1.900e-01 8.82400e-01
1.800e-01 8.27100e-01
1.700e-01 7.65500e-01
1.600e-01 6.92100e-01
1.500e-01 6.00800e-01
1.400e-01 5.00600e-01
1.300e-01 4.01800e-01
1.200e-01 3.02900e-01
1.100e-01 2.16200e-01
1.000e-01 1.36400e-01
9.000e-02 7.82000e-02
8.000e-02 4.35000e-02
7.000e-02 1.95000e-02
6.000e-02 7.60000e-03
5.000e-02 2.46002e-03
4.000e-02 7.40708e-04
3.000e-02 3.09945e-04
2.000e-02 7.30028e-05
1.000e-02 1.93102e-05
0.000e+00 8.85093e-06
};

\addplot[
    color=Paired-3,
    mark=x,
    thick,
    mark size=3,
    mark repeat=3,mark phase=3,
]
table {
3.000e-01 9.99500e-01
2.900e-01 9.98800e-01
2.800e-01 9.97600e-01
2.700e-01 9.95600e-01
2.600e-01 9.91500e-01
2.500e-01 9.82500e-01
2.400e-01 9.67800e-01
2.300e-01 9.44800e-01
2.200e-01 9.15800e-01
2.100e-01 8.68800e-01
2.000e-01 8.12000e-01
1.900e-01 7.39300e-01
1.800e-01 6.50200e-01
1.700e-01 5.46600e-01
1.600e-01 4.38200e-01
1.500e-01 3.38900e-01
1.400e-01 2.49900e-01
1.300e-01 1.75300e-01
1.200e-01 1.11800e-01
1.100e-01 6.76000e-02
1.000e-01 4.02000e-02
9.000e-02 2.04000e-02
8.000e-02 1.11000e-02
7.000e-02 5.80000e-03
6.000e-02 2.46378e-03
5.000e-02 9.13977e-04
4.000e-02 3.39259e-04
3.000e-02 1.61704e-04
2.000e-02 6.72474e-05
1.000e-02 1.78120e-05
0.000e+00 8.85093e-06
};

\addplot[
    color=Paired-5,
    mark=o,
    thick,
    mark size=3,
    mark repeat=3,mark phase=1,
]
table {
3.000e-01 9.99000e-01
2.900e-01 9.98400e-01
2.800e-01 9.97300e-01
2.700e-01 9.95800e-01
2.600e-01 9.92400e-01
2.500e-01 9.84400e-01
2.400e-01 9.78300e-01
2.300e-01 9.66500e-01
2.200e-01 9.46100e-01
2.100e-01 9.15800e-01
2.000e-01 8.77800e-01
1.900e-01 8.28000e-01
1.800e-01 7.59000e-01
1.700e-01 6.82600e-01
1.600e-01 5.97700e-01
1.500e-01 4.98300e-01
1.400e-01 3.88000e-01
1.300e-01 2.86800e-01
1.200e-01 2.05500e-01
1.100e-01 1.31200e-01
1.000e-01 7.02000e-02
9.000e-02 3.69000e-02
8.000e-02 1.69000e-02
7.000e-02 7.50000e-03
6.000e-02 1.83123e-03
5.000e-02 5.30667e-04
4.000e-02 1.10124e-04
3.000e-02 2.52066e-05
2.000e-02 7.11382e-06
1.000e-02 3.05045e-06
0.000e+00 0.00000e+00
};

\addplot[
    color=Paired-5,
    mark=x,
    thick,
    mark size=3,
    mark repeat=3,mark phase=3,
]
table {
3.000e-01 9.98900e-01
2.900e-01 9.97800e-01
2.800e-01 9.96000e-01
2.700e-01 9.92700e-01
2.600e-01 9.86900e-01
2.500e-01 9.72900e-01
2.400e-01 9.52200e-01
2.300e-01 9.21500e-01
2.200e-01 8.79500e-01
2.100e-01 8.18900e-01
2.000e-01 7.43400e-01
1.900e-01 6.53700e-01
1.800e-01 5.44200e-01
1.700e-01 4.28300e-01
1.600e-01 3.18400e-01
1.500e-01 2.22700e-01
1.400e-01 1.43500e-01
1.300e-01 8.37000e-02
1.200e-01 4.27000e-02
1.100e-01 2.10000e-02
1.000e-01 9.40000e-03
9.000e-02 3.35143e-03
8.000e-02 1.35888e-03
7.000e-02 4.74532e-04
6.000e-02 1.19331e-04
5.000e-02 5.44598e-05
4.000e-02 1.45456e-05
3.000e-02 1.10169e-05
2.000e-02 6.42566e-06
1.000e-02 3.05045e-06
};

\coordinate (bot) at (rel axis cs:1,0);

\nextgroupplot[xlabel={Channel SNR (dB)},ylabel={Average Iterations}, width = .62\columnwidth, yshift=+1.5cm, ymin=1, ymax=5e6]
\addplot[
    color=Paired-1,
    mark=o,
    thick,
    mark size=3,
    mark repeat=3,mark phase=1,
]
table {
6.000e+00 2.38e+05
6.250e+00 2.16e+05
6.500e+00 1.94e+05
6.750e+00 1.70e+05
7.000e+00 1.50e+05
7.250e+00 1.30e+05
7.500e+00 1.13e+05
7.750e+00 9.54e+04
8.000e+00 8.49e+04
8.250e+00 7.19e+04
8.500e+00 6.45e+04
8.750e+00 5.48e+04
9.000e+00 5.09e+04
9.250e+00 4.47e+04
9.500e+00 4.24e+04
9.750e+00 3.84e+04
1.000e+01 3.72e+04
1.025e+01 3.48e+04
1.050e+01 3.45e+04
1.075e+01 3.26e+04
1.100e+01 3.29e+04
1.125e+01 3.15e+04
1.150e+01 3.22e+04
1.175e+01 3.12e+04
1.200e+01 3.19e+04
};

\addplot[
    color=Paired-1,
    mark=x,
    thick,
    mark size=3,
    mark repeat=3,mark phase=3,
]
table {
6.000e+00 1.25e+05
6.250e+00 9.99e+04
6.500e+00 7.76e+04
6.750e+00 5.75e+04
7.000e+00 4.13e+04
7.250e+00 2.93e+04
7.500e+00 1.90e+04
7.750e+00 1.22e+04
8.000e+00 7.43e+03
8.250e+00 4.22e+03
8.500e+00 2.48e+03
8.750e+00 1.33e+03
9.000e+00 6.82e+02
9.250e+00 3.55e+02
9.500e+00 1.74e+02
9.750e+00 8.68e+01
1.000e+01 4.44e+01
1.025e+01 2.27e+01
1.050e+01 1.27e+01
1.075e+01 6.84e+00
1.100e+01 4.09e+00
1.125e+01 2.69e+00
1.150e+01 1.94e+00
1.175e+01 1.54e+00
1.200e+01 1.31e+00
};

\addplot[
    color=Paired-3,
    mark=o,
    thick,
    mark size=3,
    mark repeat=3,mark phase=1,
]
table {
6.000e+00 3.10e+05
6.250e+00 3.01e+05
6.500e+00 2.91e+05
6.750e+00 2.79e+05
7.000e+00 2.68e+05
7.250e+00 2.57e+05
7.500e+00 2.46e+05
7.750e+00 2.36e+05
8.000e+00 2.28e+05
8.250e+00 2.19e+05
8.500e+00 2.12e+05
8.750e+00 2.06e+05
9.000e+00 2.00e+05
9.250e+00 1.95e+05
9.500e+00 1.92e+05
9.750e+00 1.88e+05
1.000e+01 1.86e+05
1.025e+01 1.84e+05
1.050e+01 1.82e+05
1.075e+01 1.81e+05
1.100e+01 1.80e+05
1.125e+01 1.79e+05
1.150e+01 1.79e+05
1.175e+01 1.79e+05
1.200e+01 1.79e+05
};

\addplot[
    color=Paired-3,
    mark=x,
    thick,
    mark size=3,
    mark repeat=3,mark phase=3,
]
table {
6.000e+00 5.43e+04
6.250e+00 4.38e+04
6.500e+00 3.47e+04
6.750e+00 2.55e+04
7.000e+00 1.89e+04
7.250e+00 1.40e+04
7.500e+00 9.44e+03
7.750e+00 6.12e+03
8.000e+00 3.88e+03
8.250e+00 2.26e+03
8.500e+00 1.35e+03
8.750e+00 7.02e+02
9.000e+00 3.75e+02
9.250e+00 2.26e+02
9.500e+00 1.21e+02
9.750e+00 5.78e+01
1.000e+01 3.09e+01
1.025e+01 1.67e+01
1.050e+01 9.27e+00
1.075e+01 6.75e+00
1.100e+01 3.39e+00
1.125e+01 2.37e+00
1.150e+01 1.81e+00
1.175e+01 1.48e+00
1.200e+01 1.28e+00
};
\label{gp:gp_e5}

\addplot[
    color=Paired-5,
    mark=o,
    thick,
    mark size=3,
    mark repeat=3,mark phase=1,
]
table {
6.000e+00 3.39e+05
6.250e+00 3.38e+05
6.500e+00 3.37e+05
6.750e+00 3.35e+05
7.000e+00 3.34e+05
7.250e+00 3.33e+05
7.500e+00 3.31e+05
7.750e+00 3.30e+05
8.000e+00 3.29e+05
8.250e+00 3.27e+05
8.500e+00 3.26e+05
8.750e+00 3.25e+05
9.000e+00 3.23e+05
9.250e+00 3.22e+05
9.500e+00 3.21e+05
9.750e+00 3.20e+05
1.000e+01 3.20e+05
1.025e+01 3.19e+05
1.050e+01 3.19e+05
1.075e+01 3.19e+05
1.100e+01 3.19e+05
1.125e+01 3.18e+05
1.150e+01 3.18e+05
1.175e+01 3.18e+05
1.200e+01 3.18e+05
};

\addplot[
    color=Paired-5,
    mark=x,
    thick,
    mark size=3,
    mark repeat=3,mark phase=3,
]
table {
6.000e+00 4.72e+03
6.250e+00 3.92e+03
6.500e+00 3.11e+03
6.750e+00 2.53e+03
7.000e+00 2.08e+03
7.250e+00 1.65e+03
7.500e+00 1.09e+03
7.750e+00 8.25e+02
8.000e+00 5.79e+02
8.250e+00 4.08e+02
8.500e+00 2.50e+02
8.750e+00 1.46e+02
9.000e+00 7.88e+01
9.250e+00 4.40e+01
9.500e+00 2.63e+01
9.750e+00 1.60e+01
1.000e+01 9.85e+00
1.025e+01 6.62e+00
1.050e+01 4.28e+00
1.075e+01 2.99e+00
1.100e+01 2.32e+00
1.125e+01 1.78e+00
1.150e+01 1.40e+00
1.175e+01 1.27e+00
1.200e+01 1.20e+00
};

\coordinate (bot) at (rel axis cs:2,0);


\nextgroupplot[xlabel={Bit Jamming Probability ($\epsilon$)},yticklabels={,,}, width = .62\columnwidth, yshift=+1.5cm, ymin=1, ymax=5e6,x dir=reverse,
xticklabel style={
        rotate=45,
        /pgf/number format/fixed,
        /pgf/number format/precision=2
}]
\addplot[
    color=Paired-1,
    mark=o,
    thick,
    mark size=3,
    mark repeat=3,mark phase=1,
]
table {
3.000e-01 3.43e+05
2.900e-01 3.42e+05
2.800e-01 3.43e+05
2.700e-01 3.42e+05
2.600e-01 3.42e+05
2.500e-01 3.42e+05
2.400e-01 3.42e+05
2.300e-01 3.42e+05
2.200e-01 3.42e+05
2.100e-01 3.42e+05
2.000e-01 3.43e+05
1.900e-01 3.43e+05
1.800e-01 3.42e+05
1.700e-01 3.42e+05
1.600e-01 3.43e+05
1.500e-01 3.42e+05
1.400e-01 3.42e+05
1.300e-01 3.42e+05
1.200e-01 3.41e+05
1.100e-01 3.41e+05
1.000e-01 3.39e+05
9.000e-02 3.36e+05
8.000e-02 3.34e+05
7.000e-02 3.28e+05
6.000e-02 3.22e+05
5.000e-02 3.10e+05
4.000e-02 2.94e+05
3.000e-02 2.69e+05
2.000e-02 2.38e+05
1.000e-02 2.00e+05
0.000e+00 1.58e+05
};

\addplot[
    color=Paired-1,
    mark=x,
    thick,
    mark size=3,
    mark repeat=3,mark phase=3,
]
table {
3.000e-01 1.00e+00
2.900e-01 1.00e+00
2.800e-01 1.02e+00
2.700e-01 1.02e+00
2.600e-01 1.05e+00
2.500e-01 1.15e+00
2.400e-01 1.36e+00
2.300e-01 1.84e+00
2.200e-01 2.81e+00
2.100e-01 4.41e+00
2.000e-01 9.53e+00
1.900e-01 1.45e+01
1.800e-01 2.90e+01
1.700e-01 5.45e+01
1.600e-01 1.27e+02
1.500e-01 2.62e+02
1.400e-01 4.60e+02
1.300e-01 8.82e+02
1.200e-01 1.57e+03
1.100e-01 2.86e+03
1.000e-01 4.72e+03
9.000e-02 8.23e+03
8.000e-02 1.38e+04
7.000e-02 2.30e+04
6.000e-02 3.63e+04
5.000e-02 5.43e+04
4.000e-02 7.70e+04
3.000e-02 1.01e+05
2.000e-02 1.25e+05
1.000e-02 1.43e+05
0.000e+00 1.57e+05
};

\addplot[
    color=Paired-3,
    mark=o,
    thick,
    mark size=3,
    mark repeat=3,mark phase=1,
]
table {
3.000e-01 3.43e+05
2.900e-01 3.42e+05
2.800e-01 3.42e+05
2.700e-01 3.42e+05
2.600e-01 3.42e+05
2.500e-01 3.42e+05
2.400e-01 3.42e+05
2.300e-01 3.43e+05
2.200e-01 3.42e+05
2.100e-01 3.42e+05
2.000e-01 3.43e+05
1.900e-01 3.42e+05
1.800e-01 3.42e+05
1.700e-01 3.41e+05
1.600e-01 3.42e+05
1.500e-01 3.42e+05
1.400e-01 3.40e+05
1.300e-01 3.39e+05
1.200e-01 3.36e+05
1.100e-01 3.33e+05
1.000e-01 3.29e+05
9.000e-02 3.19e+05
8.000e-02 3.07e+05
7.000e-02 2.87e+05
6.000e-02 2.63e+05
5.000e-02 2.27e+05
4.000e-02 1.87e+05
3.000e-02 1.35e+05
2.000e-02 8.49e+04
1.000e-02 3.90e+04
0.000e+00 9.42e+03
};

\addplot[
    color=Paired-3,
    mark=x,
    thick,
    mark size=3,
    mark repeat=3,mark phase=3,
]
table {
3.000e-01 1.00e+00
2.900e-01 1.00e+00
2.800e-01 1.01e+00
2.700e-01 1.02e+00
2.600e-01 1.04e+00
2.500e-01 1.10e+00
2.400e-01 1.23e+00
2.300e-01 1.34e+00
2.200e-01 1.69e+00
2.100e-01 2.44e+00
2.000e-01 3.95e+00
1.900e-01 8.84e+00
1.800e-01 1.31e+01
1.700e-01 1.89e+01
1.600e-01 3.20e+01
1.500e-01 7.22e+01
1.400e-01 1.04e+02
1.300e-01 1.45e+02
1.200e-01 2.55e+02
1.100e-01 4.02e+02
1.000e-01 5.79e+02
9.000e-02 8.97e+02
8.000e-02 1.30e+03
7.000e-02 1.89e+03
6.000e-02 2.74e+03
5.000e-02 3.88e+03
4.000e-02 4.97e+03
3.000e-02 6.38e+03
2.000e-02 7.43e+03
1.000e-02 8.60e+03
0.000e+00 9.35e+03
};

\addplot[
    color=Paired-5,
    mark=o,
    thick,
    mark size=3,
    mark repeat=3,mark phase=1,
]
table {
3.000e-01 3.43e+05
2.900e-01 3.42e+05
2.800e-01 3.42e+05
2.700e-01 3.42e+05
2.600e-01 3.42e+05
2.500e-01 3.42e+05
2.400e-01 3.42e+05
2.300e-01 3.42e+05
2.200e-01 3.42e+05
2.100e-01 3.43e+05
2.000e-01 3.42e+05
1.900e-01 3.42e+05
1.800e-01 3.42e+05
1.700e-01 3.41e+05
1.600e-01 3.42e+05
1.500e-01 3.41e+05
1.400e-01 3.38e+05
1.300e-01 3.36e+05
1.200e-01 3.31e+05
1.100e-01 3.27e+05
1.000e-01 3.18e+05
9.000e-02 3.05e+05
8.000e-02 2.85e+05
7.000e-02 2.58e+05
6.000e-02 2.23e+05
5.000e-02 1.78e+05
4.000e-02 1.28e+05
3.000e-02 7.44e+04
2.000e-02 3.19e+04
1.000e-02 6.54e+03
0.000e+00 1.32e+00
};

\addplot[
    color=Paired-5,
    mark=x,
    thick,
    mark size=3,
    mark repeat=3,mark phase=3,
]
table {
3.000e-01 1.00e+00
2.900e-01 1.00e+00
2.800e-01 1.00e+00
2.700e-01 1.00e+00
2.600e-01 1.00e+00
2.500e-01 1.00e+00
2.400e-01 1.00e+00
2.300e-01 1.00e+00
2.200e-01 1.00e+00
2.100e-01 1.00e+00
2.000e-01 1.01e+00
1.900e-01 1.01e+00
1.800e-01 1.02e+00
1.700e-01 1.04e+00
1.600e-01 1.06e+00
1.500e-01 1.06e+00
1.400e-01 1.09e+00
1.300e-01 1.12e+00
1.200e-01 1.16e+00
1.100e-01 1.18e+00
1.000e-01 1.20e+00
9.000e-02 1.26e+00
8.000e-02 1.24e+00
7.000e-02 1.25e+00
6.000e-02 1.25e+00
5.000e-02 1.28e+00
4.000e-02 1.29e+00
3.000e-02 1.30e+00
2.000e-02 1.31e+00
1.000e-02 1.32e+00
0.000e+00 1.32e+00
};

\coordinate (top) at (rel axis cs:0,1);

\nextgroupplot[xlabel={Channel SNR (dB)}, yticklabels={,,},width = .62\columnwidth, yshift=+1.5cm, ymin=1, ymax=5e6]
\addplot[
    color=Paired-1,
    mark=o,
    thick,
    mark size=3,
    mark repeat=3,mark phase=1,
]
table {
6.000e+00 1.83e+05
6.250e+00 9.42e+04
6.500e+00 4.36e+04
6.750e+00 1.87e+04
7.000e+00 1.01e+04
7.250e+00 4.58e+03
7.500e+00 2.12e+03
7.750e+00 1.01e+03
8.000e+00 5.31e+02
8.250e+00 2.88e+02
8.500e+00 1.76e+02
8.750e+00 1.30e+02
9.000e+00 9.95e+01
9.250e+00 7.25e+01
9.500e+00 5.92e+01
9.750e+00 5.52e+01
1.000e+01 6.31e+01
1.025e+01 6.02e+01
1.050e+01 4.82e+01
1.075e+01 5.71e+01
1.100e+01 5.56e+01
1.125e+01 4.32e+01
1.150e+01 4.14e+01
1.175e+01 5.31e+01
1.200e+01 3.95e+01
};

\addplot[
    color=Paired-1,
    mark=x,
    thick,
    mark size=3,
    mark repeat=3,mark phase=3,
]
table {
6.000e+00 1.29e+05
6.250e+00 6.23e+04
6.500e+00 2.83e+04
6.750e+00 1.21e+04
7.000e+00 5.74e+03
7.250e+00 2.28e+03
7.500e+00 9.54e+02
7.750e+00 4.64e+02
8.000e+00 2.25e+02
8.250e+00 1.05e+02
8.500e+00 5.77e+01
8.750e+00 3.34e+01
9.000e+00 2.59e+01
9.250e+00 2.26e+01
9.500e+00 1.80e+01
9.750e+00 1.80e+01
1.000e+01 1.81e+01
1.025e+01 1.85e+01
1.050e+01 1.88e+01
1.075e+01 1.91e+01
1.100e+01 1.96e+01
1.125e+01 1.97e+01
1.150e+01 1.99e+01
1.175e+01 2.01e+01
1.200e+01 2.05e+01
};

\addplot[
    color=Paired-3,
    mark=o,
    thick,
    mark size=3,
    mark repeat=3,mark phase=1,
]
table {
6.000e+00 6.76e+05
6.250e+00 4.47e+05
6.500e+00 2.81e+05
6.750e+00 1.80e+05
7.000e+00 1.15e+05
7.250e+00 7.50e+04
7.500e+00 4.67e+04
7.750e+00 2.93e+04
8.000e+00 2.16e+04
8.250e+00 1.56e+04
8.500e+00 1.14e+04
8.750e+00 9.37e+03
9.000e+00 7.66e+03
9.250e+00 6.66e+03
9.500e+00 5.95e+03
9.750e+00 5.20e+03
1.000e+01 4.81e+03
1.025e+01 4.43e+03
1.050e+01 4.18e+03
1.075e+01 4.11e+03
1.100e+01 3.93e+03
1.125e+01 3.98e+03
1.150e+01 3.89e+03
1.175e+01 3.83e+03
1.200e+01 3.79e+03
};

\addplot[
    color=Paired-3,
    mark=x,
    thick,
    mark size=3,
    mark repeat=3,mark phase=3,
]
table {
6.000e+00 1.87e+05
6.250e+00 1.05e+05
6.500e+00 5.42e+04
6.750e+00 2.75e+04
7.000e+00 1.36e+04
7.250e+00 6.35e+03
7.500e+00 2.85e+03
7.750e+00 1.22e+03
8.000e+00 6.97e+02
8.250e+00 3.09e+02
8.500e+00 2.27e+02
8.750e+00 1.01e+02
9.000e+00 1.31e+02
9.250e+00 5.56e+01
9.500e+00 1.09e+02
9.750e+00 4.43e+01
1.000e+01 9.90e+01
1.025e+01 4.11e+01
1.050e+01 8.73e+01
1.075e+01 3.82e+01
1.100e+01 7.18e+01
1.125e+01 3.79e+01
1.150e+01 7.15e+01
1.175e+01 3.75e+01
1.200e+01 6.92e+01
};

\addplot[
    color=Paired-5,
    mark=o,
    thick,
    mark size=3,
    mark repeat=3,mark phase=1,
]
table {
6.000e+00 2.61e+06
6.250e+00 2.26e+06
6.500e+00 1.95e+06
6.750e+00 1.65e+06
7.000e+00 1.43e+06
7.250e+00 1.23e+06
7.500e+00 1.06e+06
7.750e+00 9.29e+05
8.000e+00 8.25e+05
8.250e+00 7.37e+05
8.500e+00 6.70e+05
8.750e+00 6.18e+05
9.000e+00 5.81e+05
9.250e+00 5.46e+05
9.500e+00 5.25e+05
9.750e+00 5.09e+05
1.000e+01 4.96e+05
1.025e+01 4.84e+05
1.050e+01 4.78e+05
1.075e+01 4.74e+05
1.100e+01 4.70e+05
1.125e+01 4.66e+05
1.150e+01 4.63e+05
1.175e+01 4.62e+05
1.200e+01 4.61e+05
};

\addplot[
    color=Paired-5,
    mark=x,
    thick,
    mark size=3,
    mark repeat=3,mark phase=3,
]
table {
6.000e+00 1.01e+05
6.250e+00 6.73e+04
6.500e+00 4.13e+04
6.750e+00 2.55e+04
7.000e+00 1.54e+04
7.250e+00 9.31e+03
7.500e+00 4.64e+03
7.750e+00 2.44e+03
8.000e+00 1.54e+03
8.250e+00 8.52e+02
8.500e+00 4.79e+02
8.750e+00 2.80e+02
9.000e+00 1.64e+02
9.250e+00 1.05e+02
9.500e+00 7.19e+01
9.750e+00 5.16e+01
1.000e+01 4.01e+01
1.025e+01 3.33e+01
1.050e+01 2.93e+01
1.075e+01 2.74e+01
1.100e+01 2.53e+01
1.125e+01 2.69e+01
1.150e+01 2.08e+01
1.175e+01 2.66e+01
1.200e+01 2.20e+01
};

\coordinate (bot) at (rel axis cs:1,1);

\nextgroupplot[xlabel={Bit Jamming Probability ($\epsilon$)},yticklabels={,,}, width = .62\columnwidth, yshift=+1.5cm, ymin=1, ymax=5e6,x dir=reverse,
xticklabel style={
        rotate=45,
        /pgf/number format/fixed,
        /pgf/number format/precision=2
}]
\addplot[
    color=Paired-1,
    mark=o,
    thick,
    mark size=3,
    mark repeat=3,mark phase=1,
]
table {
3.000e-01 4.93e+06
2.900e-01 4.99e+06
2.800e-01 4.93e+06
2.700e-01 4.97e+06
2.600e-01 4.96e+06
2.500e-01 4.86e+06
2.400e-01 4.92e+06
2.300e-01 4.89e+06
2.200e-01 4.92e+06
2.100e-01 4.85e+06
2.000e-01 4.86e+06
1.900e-01 4.78e+06
1.800e-01 4.72e+06
1.700e-01 4.59e+06
1.600e-01 4.50e+06
1.500e-01 4.29e+06
1.400e-01 4.07e+06
1.300e-01 3.80e+06
1.200e-01 3.43e+06
1.100e-01 3.03e+06
1.000e-01 2.61e+06
9.000e-02 2.17e+06
8.000e-02 1.72e+06
7.000e-02 1.31e+06
6.000e-02 9.55e+05
5.000e-02 6.70e+05
4.000e-02 4.41e+05
3.000e-02 2.85e+05
2.000e-02 1.83e+05
1.000e-02 1.15e+05
0.000e+00 7.27e+04
};

\addplot[
    color=Paired-1,
    mark=x,
    thick,
    mark size=3,
    mark repeat=3,mark phase=3,
]
table {
3.000e-01 1.62e+00
2.900e-01 2.60e+00
2.800e-01 4.85e+00
2.700e-01 9.75e+00
2.600e-01 2.10e+01
2.500e-01 4.31e+01
2.400e-01 8.92e+01
2.300e-01 1.82e+02
2.200e-01 3.61e+02
2.100e-01 6.80e+02
2.000e-01 1.20e+03
1.900e-01 2.02e+03
1.800e-01 3.77e+03
1.700e-01 6.65e+03
1.600e-01 1.09e+04
1.500e-01 1.84e+04
1.400e-01 2.86e+04
1.300e-01 4.04e+04
1.200e-01 5.88e+04
1.100e-01 7.64e+04
1.000e-01 1.01e+05
9.000e-02 1.27e+05
8.000e-02 1.48e+05
7.000e-02 1.67e+05
6.000e-02 1.81e+05
5.000e-02 1.87e+05
4.000e-02 1.75e+05
3.000e-02 1.56e+05
2.000e-02 1.29e+05
1.000e-02 9.90e+04
0.000e+00 7.26e+04
};

\addplot[
    color=Paired-3,
    mark=o,
    thick,
    mark size=3,
    mark repeat=3,mark phase=1,
]
table {
3.000e-01 4.97e+06
2.900e-01 4.97e+06
2.800e-01 4.93e+06
2.700e-01 4.90e+06
2.600e-01 4.95e+06
2.500e-01 4.92e+06
2.400e-01 4.90e+06
2.300e-01 4.83e+06
2.200e-01 4.84e+06
2.100e-01 4.74e+06
2.000e-01 4.62e+06
1.900e-01 4.47e+06
1.800e-01 4.24e+06
1.700e-01 3.90e+06
1.600e-01 3.62e+06
1.500e-01 3.18e+06
1.400e-01 2.74e+06
1.300e-01 2.22e+06
1.200e-01 1.72e+06
1.100e-01 1.27e+06
1.000e-01 8.25e+05
9.000e-02 5.13e+05
8.000e-02 2.99e+05
7.000e-02 1.46e+05
6.000e-02 5.81e+04
5.000e-02 2.13e+04
4.000e-02 6.27e+03
3.000e-02 1.79e+03
2.000e-02 5.31e+02
1.000e-02 1.81e+02
0.000e+00 7.81e+01
};

\addplot[
    color=Paired-3,
    mark=x,
    thick,
    mark size=3,
    mark repeat=3,mark phase=3,
]
table {
3.000e-01 1.16e+00
2.900e-01 1.39e+00
2.800e-01 1.84e+00
2.700e-01 2.82e+00
2.600e-01 4.67e+00
2.500e-01 8.27e+00
2.400e-01 1.53e+01
2.300e-01 2.75e+01
2.200e-01 4.82e+01
2.100e-01 8.21e+01
2.000e-01 1.34e+02
1.900e-01 2.10e+02
1.800e-01 3.17e+02
1.700e-01 4.60e+02
1.600e-01 7.62e+02
1.500e-01 9.05e+02
1.400e-01 1.22e+03
1.300e-01 1.25e+03
1.200e-01 1.31e+03
1.100e-01 1.30e+03
1.000e-01 1.30e+03
9.000e-02 1.19e+03
8.000e-02 1.03e+03
7.000e-02 9.33e+02
6.000e-02 6.87e+02
5.000e-02 5.08e+02
4.000e-02 3.93e+02
3.000e-02 3.18e+02
2.000e-02 2.45e+02
1.000e-02 1.32e+02
0.000e+00 7.81e+01
};

\addplot[
    color=Paired-5,
    mark=o,
    thick,
    mark size=3,
    mark repeat=3,mark phase=1,
]
table {
3.000e-01 4.94e+06
2.900e-01 4.96e+06
2.800e-01 4.97e+06
2.700e-01 4.92e+06
2.600e-01 4.94e+06
2.500e-01 4.94e+06
2.400e-01 4.82e+06
2.300e-01 4.81e+06
2.200e-01 4.77e+06
2.100e-01 4.62e+06
2.000e-01 4.46e+06
1.900e-01 4.25e+06
1.800e-01 3.93e+06
1.700e-01 3.54e+06
1.600e-01 3.19e+06
1.500e-01 2.67e+06
1.400e-01 2.19e+06
1.300e-01 1.65e+06
1.200e-01 1.18e+06
1.100e-01 7.85e+05
1.000e-01 4.61e+05
9.000e-02 2.50e+05
8.000e-02 1.21e+05
7.000e-02 5.24e+04
6.000e-02 1.47e+04
5.000e-02 3.79e+03
4.000e-02 7.58e+02
3.000e-02 1.78e+02
2.000e-02 5.33e+01
1.000e-02 2.01e+01
0.000e+00 1.01e+00
};

\addplot[
    color=Paired-5,
    mark=x,
    thick,
    mark size=3,
    mark repeat=3,mark phase=3,
]
table {
3.000e-01 1.00e+00
2.900e-01 1.00e+00
2.800e-01 1.00e+00
2.700e-01 1.01e+00
2.600e-01 1.01e+00
2.500e-01 1.03e+00
2.400e-01 1.06e+00
2.300e-01 1.12e+00
2.200e-01 1.18e+00
2.100e-01 1.29e+00
2.000e-01 1.50e+00
1.900e-01 1.64e+00
1.800e-01 2.25e+00
1.700e-01 3.11e+00
1.600e-01 3.31e+00
1.500e-01 5.92e+00
1.400e-01 7.30e+00
1.300e-01 9.75e+00
1.200e-01 1.23e+01
1.100e-01 1.77e+01
1.000e-01 1.97e+01
9.000e-02 2.80e+01
8.000e-02 3.90e+01
7.000e-02 3.92e+01
6.000e-02 3.87e+01
5.000e-02 4.03e+01
4.000e-02 4.86e+01
3.000e-02 3.86e+01
2.000e-02 2.06e+01
1.000e-02 1.39e+01
0.000e+00 1.01e+00
};

\coordinate (bot) at (rel axis cs:2,1);

\end{groupplot}

    \node (A) at (0.30,0.35) {\Large{(a)}};
    \node (B) at (4.55,0.35) {\Large{(b)}};
    \node (C) at (8.70,0.35) {\Large{(c)}};
    \node (D) at (12.95,0.35) {\Large{(d)}};
    
    \node (E) at (3.60,-0.82) {\Large{(e)}};
    \node (F) at (7.80,-0.82) {\Large{(f)}};
    \node (G) at (12.00,-0.82) {\Large{(g)}};
    \node (H) at (16.20,-0.82) {\Large{(h)}};

    \path (top|-current bounding box.north) -- 
      coordinate(legendpos)
      (bot|-current bounding box.north);
    \matrix[
        matrix of nodes,
        anchor=south,
        draw,
        inner sep=0.2em,
        draw
      ]at([yshift=+.3ex,xshift=-65.75ex]legendpos)
      {
        \ref{gp:g_e2}& \scriptsize GRAND ($\epsilon\text{:}0.02$) \\
        \ref{gp:g_e5}& \scriptsize GRAND ($\epsilon\text{:}0.05$) \\
        \ref{gp:g_e10}& \scriptsize GRAND ($\epsilon\text{:}0.10$) \\
        \ref{gp:ge_e2}& \scriptsize GRAND-EDGE ($\epsilon\text{:}0.02$) \\
        \ref{gp:ge_e5}& \scriptsize GRAND-EDGE ($\epsilon\text{:}0.05$) \\
        \ref{gp:ge_e10}& \scriptsize GRAND-EDGE ($\epsilon\text{:}0.10$) \\};

    \path (top|-current bounding box.north) -- 
      coordinate(legendpos)
      (bot|-current bounding box.north);
    \matrix[
        matrix of nodes,
        anchor=south,
        draw,
        inner sep=0.2em,
        draw
      ]at([yshift=-15.3ex,xshift=-40.0ex]legendpos)
      {
        \ref{gp:g_e2}& \scriptsize GRAND (SNR:6dB) \\
        \ref{gp:g_e5}& \scriptsize GRAND (SNR:8dB) \\
        \ref{gp:g_e10}& \scriptsize GRAND (SNR:12dB) \\
        \ref{gp:ge_e2}& \scriptsize GRAND-EDGE (SNR:6dB) \\
        \ref{gp:ge_e5}& \scriptsize GRAND-EDGE (SNR:8dB) \\
        \ref{gp:ge_e10}& \scriptsize GRAND-EDGE (SNR:12dB) \\};
        
    \path (top|-current bounding box.north) -- 
      coordinate(legendpos)
      (bot|-current bounding box.north);
    \matrix[
        matrix of nodes,
        anchor=south,
        draw,
        inner sep=0.2em,
        draw
      ]at([yshift=-15.4ex,xshift=-13.5ex]legendpos)
      {
        \ref{gp:g_e2}& \scriptsize ORBGRAND ($\epsilon\text{:}0.02$) \\
        \ref{gp:g_e5}& \scriptsize ORBGRAND ($\epsilon\text{:}0.05$) \\
        \ref{gp:g_e10}& \scriptsize ORBGRAND ($\epsilon\text{:}0.10$) \\
        \ref{gp:ge_e2}& \scriptsize ORBGRAND-EDGE ($\epsilon\text{:}0.02$) \\
        \ref{gp:ge_e5}& \scriptsize ORBGRAND-EDGE ($\epsilon\text{:}0.05$) \\
        \ref{gp:ge_e10}& \scriptsize ORBGRAND-EDGE ($\epsilon\text{:}0.10$) \\};

    \path (top|-current bounding box.north) -- 
      coordinate(legendpos)
      (bot|-current bounding box.north);
    \matrix[
        matrix of nodes,
        anchor=south,
        draw,
        inner sep=0.2em,
        draw
      ]at([yshift=-15.3ex,xshift=+13.5ex]legendpos)
      {
        \ref{gp:g_e2}& \scriptsize ORBGRAND (SNR:6dB) \\
        \ref{gp:g_e5}& \scriptsize ORBGRAND (SNR:8dB) \\
        \ref{gp:g_e10}& \scriptsize ORBGRAND (SNR:12dB) \\
        \ref{gp:ge_e2}& \scriptsize ORBGRAND-EDGE (SNR:6dB) \\
        \ref{gp:ge_e5}& \scriptsize ORBGRAND-EDGE (SNR:8dB) \\
        \ref{gp:ge_e10}& \scriptsize ORBGRAND-EDGE (SNR:12dB) \\};
    
\end{tikzpicture}

%% file: figures/results_osd_p_bler.tikz
\begin{tikzpicture}[spy using outlines=
	{circle, magnification=2.0, connect spies}]
  \pgfplotsset{
    label style = {font=\fontsize{9pt}{7.2}\selectfont},
    tick label style = {font=\fontsize{7pt}{7.2}\selectfont}
  }

\begin{axis}[
	scale = 1,
    ymode=log,
    xlabel={Channel SNR (dB)}, xlabel style={yshift=0.4em},
    ylabel={BLER}, ylabel style={yshift=-0.50em},
    ymax = 2,
    grid=both,
    ymajorgrids=true,
    xmajorgrids=true,
    grid style=dashed,
    width=\columnwidth, height=6.0cm,
    thick,
    mark size=3,
    legend cell align={left},
    legend style={at={(-0.01,1.20)},anchor=west,
            font=\fontsize{8pt}{7.2}\selectfont},
    legend columns=2,
]
\addplot[
    color=Paired-1,
    mark=o,
    thick,
    mark size=3,
    mark repeat=3,mark phase=1,
]
table {
6.000e+00 1.17100e-01
6.250e+00 8.02000e-02
6.500e+00 5.41000e-02
6.750e+00 3.69000e-02
7.000e+00 2.39000e-02
7.250e+00 1.66000e-02
7.500e+00 9.80000e-03
7.750e+00 7.60000e-03
8.000e+00 4.51264e-03
8.250e+00 3.92188e-03
8.500e+00 3.07579e-03
8.750e+00 2.93513e-03
9.000e+00 1.98633e-03
9.250e+00 1.61770e-03
9.500e+00 1.55169e-03
9.750e+00 1.25575e-03
1.000e+01 1.07492e-03
1.025e+01 9.26904e-04
1.050e+01 7.96369e-04
1.075e+01 7.87240e-04
1.100e+01 7.52978e-04
1.125e+01 7.49749e-04
1.150e+01 7.21022e-04
1.175e+01 7.39098e-04
1.200e+01 7.31722e-04
};
\addlegendentry{OSD ($\epsilon=0.02$)}

\addplot[
    color=Paired-1,
    mark=x,
    thick,
    mark size=3,
    mark repeat=3,mark phase=3,
]
table {
6.000e+00 2.57000e-02
6.250e+00 1.29000e-02
6.500e+00 5.70000e-03
6.750e+00 2.50803e-03
7.000e+00 1.09572e-03
7.250e+00 5.13780e-04
7.500e+00 2.78413e-04
7.750e+00 1.19863e-04
8.000e+00 6.52846e-05
8.250e+00 2.74445e-05
8.500e+00 1.61963e-05
8.750e+00 1.07768e-05
9.000e+00 9.43975e-06
9.250e+00 7.27315e-06
9.500e+00 6.59435e-06
9.750e+00 6.59435e-06
1.000e+01 6.57762e-06
1.025e+01 6.48311e-06
1.050e+01 5.70818e-06
1.075e+01 5.52828e-06
1.100e+01 5.30363e-06
1.125e+01 5.27909e-06
1.150e+01 6.46918e-06
1.175e+01 6.42566e-06
1.200e+01 4.92720e-06
};
\addlegendentry{ORBGRAND-EDGE ($\epsilon=0.02$)}

\addplot[
    color=Paired-3,
    mark=o,
    thick,
    mark size=3,
    mark repeat=3,mark phase=1,
]
table {
6.000e+00 3.45500e-01
6.250e+00 2.83600e-01
6.500e+00 2.33400e-01
6.750e+00 1.92600e-01
7.000e+00 1.59100e-01
7.250e+00 1.32100e-01
7.500e+00 1.10900e-01
7.750e+00 9.46000e-02
8.000e+00 8.13000e-02
8.250e+00 7.21000e-02
8.500e+00 6.48000e-02
8.750e+00 5.78000e-02
9.000e+00 5.32000e-02
9.250e+00 4.86000e-02
9.500e+00 4.56000e-02
9.750e+00 4.30000e-02
1.000e+01 4.21000e-02
1.025e+01 4.12000e-02
1.050e+01 3.99000e-02
1.075e+01 3.93000e-02
1.100e+01 3.87000e-02
1.125e+01 3.82000e-02
1.150e+01 3.81000e-02
1.175e+01 3.80000e-02
1.200e+01 3.80000e-02
};
\addlegendentry{OSD ($\epsilon=0.05$)}

\addplot[
    color=Paired-3,
    mark=x,
    thick,
    mark size=3,
    mark repeat=3,mark phase=3,
]
table {
6.000e+00 9.13000e-02
6.250e+00 5.41000e-02
6.500e+00 3.34000e-02
6.750e+00 1.85000e-02
7.000e+00 9.50000e-03
7.250e+00 5.50000e-03
7.500e+00 2.60974e-03
7.750e+00 1.50376e-03
8.000e+00 9.13977e-04
8.250e+00 5.28290e-04
8.500e+00 2.78152e-04
8.750e+00 1.88534e-04
9.000e+00 1.28064e-04
9.250e+00 9.02122e-05
9.500e+00 7.96213e-05
9.750e+00 5.99304e-05
1.000e+01 6.70632e-05
1.025e+01 6.44366e-05
1.050e+01 5.06457e-05
1.075e+01 5.57468e-05
1.100e+01 4.82110e-05
1.125e+01 5.47406e-05
1.150e+01 4.82110e-05
1.175e+01 5.47406e-05
1.200e+01 4.60444e-05
};
\addlegendentry{ORBGRAND-EDGE ($\epsilon=0.05$)}

\addplot[
    color=Paired-5,
    mark=o,
    thick,
    mark size=3,
    mark repeat=3,mark phase=1,
]
table {
6.000e+00 7.53100e-01
6.250e+00 7.10700e-01
6.500e+00 6.76600e-01
6.750e+00 6.36600e-01
7.000e+00 6.01300e-01
7.250e+00 5.70200e-01
7.500e+00 5.39700e-01
7.750e+00 5.13800e-01
8.000e+00 4.89700e-01
8.250e+00 4.70800e-01
8.500e+00 4.54300e-01
8.750e+00 4.41200e-01
9.000e+00 4.29400e-01
9.250e+00 4.20000e-01
9.500e+00 4.11100e-01
9.750e+00 4.05600e-01
1.000e+01 4.01100e-01
1.025e+01 3.97900e-01
1.050e+01 3.95000e-01
1.075e+01 3.91900e-01
1.100e+01 3.90100e-01
1.125e+01 3.88700e-01
1.150e+01 3.87700e-01
1.175e+01 3.87300e-01
1.200e+01 3.87100e-01
};
\addlegendentry{OSD ($\epsilon=0.10$)}

\addplot[
    color=Paired-5,
    mark=x,
    thick,
    mark size=3,
    mark repeat=3,mark phase=3,
]
table {
6.000e+00 3.72900e-01
6.250e+00 2.95600e-01
6.500e+00 2.33500e-01
6.750e+00 1.79300e-01
7.000e+00 1.34800e-01
7.250e+00 1.00100e-01
7.500e+00 7.32000e-02
7.750e+00 5.39000e-02
8.000e+00 4.02000e-02
8.250e+00 3.01000e-02
8.500e+00 2.28000e-02
8.750e+00 1.90000e-02
9.000e+00 1.60000e-02
9.250e+00 1.35000e-02
9.500e+00 1.19000e-02
9.750e+00 1.12000e-02
1.000e+01 1.05000e-02
1.025e+01 1.00000e-02
1.050e+01 9.90000e-03
1.075e+01 9.80000e-03
1.100e+01 9.70000e-03
1.125e+01 9.50000e-03
1.150e+01 9.40000e-03
1.175e+01 9.40000e-03
1.200e+01 9.40000e-03
};
\addlegendentry{ORBGRAND-EDGE ($\epsilon=0.10$)}

\end{axis}

\end{tikzpicture}